\documentclass[%
 reprint,
superscriptaddress,
showkeys,
 amsmath,amssymb,
 aps,
pre,
onecolumn
]{revtex4-2}
\usepackage[a4paper, total={6in, 8in}]{geometry}
\usepackage{graphicx}
\usepackage[caption=false]{subfig} 
\usepackage{dcolumn}
\usepackage{bm}
\usepackage[mathscr]{euscript}  
\usepackage{xcolor}  

\usepackage{tcolorbox}
\usepackage{float}
\usepackage{appendix}
\begin{document} 

\preprint{APS/123-QED}
\title{ Entropy production of resetting processes }

\author{Francesco Mori}
\affiliation{LPTMS, CNRS, Univ. Paris-Sud, Universit\'{e} Paris-Saclay, 91405 Orsay, France\\}
\affiliation{Rudolf Peierls Centre for Theoretical Physics, University of Oxford, Oxford, United Kingdom\\}
\author{Kristian St\o{}levik Olsen}
\affiliation{Nordita, Royal Institute of Technology and Stockholm University,Hannes Alfvéns väg 12, 23, SE-106 91 Stockholm, Sweden\\}
\author{Supriya Krishnamurthy}
\affiliation{Department of Physics, Stockholm University, SE-10691 Stockholm, Sweden\\}

\begin{abstract} 
Stochastic systems that undergo random restarts to their initial state have been widely investigated in recent years, both theoretically and in experiments. Oftentimes, however, resetting to a fixed state is impossible due to thermal noise or other limitations. As a result, the system configuration after a resetting event is random. Here, we consider such a resetting protocol for an overdamped Brownian particle in a confining potential $V(x)$. We assume that the position of the particle is reset at a constant rate to a random location $x$, drawn from a distribution $p_R(x)$. To investigate the thermodynamic cost of resetting, we study the stochastic entropy production $S_{\rm Total}$. We derive a general expression for the average entropy production for any $V(x)$, 
and the full distribution $P(S_{\rm Total}|t)$ of the entropy production for $V(x)=0$. At late times, we show that this distribution assumes the large-deviation form $P(S_{\rm Total}|t)\sim \exp\left[-t^{2\alpha-1}\phi\left(\left(S_{\rm Total}-\langle S_{\rm Total}\rangle\right)/t^{\alpha}\right)\right]$, with $1/2<\alpha\leq 1$. We compute the rate function $\phi(z)$ and the exponent $\alpha$ for exponential and Gaussian resetting distributions. In the latter case, we find the anomalous exponent $\alpha=2/3$ and show that $\phi(z)$ has a first-order singularity at a critical value of $z$, corresponding to a real-space condensation transition.
\end{abstract}

\pacs{Valid PACS appear here} 
\maketitle

\section{Introduction}

Stochastic processes with random restarts have been extensively studied over the last decade \cite{evans2011reset,evans2020review}. In the typical setting, the resetting dynamics induces a steady state with manifest violations of detailed balance, driving the system out of equilibrium. For this reason, stochastic resetting is very interesting from both a dynamic as well as a thermodynamic perspective. \emph{Perfect resetting}, where the system is always restarted from the same state, is an example of a process with unidirectional transitions, which falls outside the normal scope of stochastic thermodynamics. Yet there are several physical processes where resetting in some form is known to play a role, such as the erasure of a bit of information under thermal fluctuations \cite{Berut2012MD,Koski2014szilard, Roldan2014symm}, or biological systems \cite{Roldan2016bio, Lisica2016back, Bressloff2020cyto, Bressloff2020intra, Genthon2022cell}. Hence, understanding the thermodynamic properties of resetting and quantifying its thermodynamic cost is a problem of general interest with applications across fields.

Perfect resetting is a unidirectional process since it has no time-reversed equivalent. The issue of how to compute the thermodynamic cost for unidirectional processes (or absolutely irreversible processes as they are called in \cite{Ueda2014_irrev}) has received some attention lately (see \cite{Busiello2020uni} and references therein). The stochastic thermodynamics of perfect resetting has been addressed for both discrete jump processes and diffusive systems in Ref.~\cite{fuchs2016stochastic}. In both cases, an average entropic contribution of resetting is identified. Taking this contribution into account, stochastic resetting systems have  been shown to satisfy integral fluctuation theorems in \cite{pal2017integral}.
In addition, work fluctuations have been calculated for a system with simultaneous particle and protocol resets \cite{gupta2020work}, though by ignoring the work required for the resetting process. Thermodynamic uncertainty relations for systems with a combination of unidirectional and bidirectional transitions, including processes with stochastic resetting, have also been studied \cite{pal2021thermodynamic}. 
However, none of the above provide a framework for calculating the distribution of entropy production for a process with instantaneous resetting.

In this paper, we revisit the issue of estimating the entropy production of a resetting process. We look at a process that mitigates the unidirectional character of resetting by restarting from a variable position picked from a distribution $p_R$ (akin to resetting with errors). However, resetting is still instantaneous. For such a process, trajectory-wise entropies can be defined as usually done in stochastic thermodynamics \cite{Seifert:2005epa}. 

We note that similar models have been studied before in the context of steady states or other dynamical properties,  beginning with an early paper on stochastic multiplicative processes with resetting \cite{Manrubia1999dis} to more recent works on resetting to random positions in the context of first passage times of Brownian processes \cite{evans2011diffusion,toledo2022first} as well as for steady-states of random walks on networks \cite{gonzalez2021diffusive}. The steady state of a Brownian particle undergoing resetting to random positions has also been studied using a renewal approach in \cite{evans2020review}. A resetting distribution that is conditioned on the position of the particle before resetting has also been considered \cite{dahlenburg2021stochastic}.

Resetting to a random position emerges naturally in experiments on colloids \cite{besga2020PRR,faisant2021_2d}, where the resetting events are usually performed by switching on and off an external potential (generated via optical tweezers). Ref.~\cite{besga2020PRR} demonstrated that a finite spread in the resetting position in $1d$ leads to a phase transition in the mean first-passage time as a function of resetting rate/period. More recently, this has been demonstrated in $2$d as well \cite{faisant2021_2d}. 
For such systems, the thermodynamic cost of resetting would be related to the work required to switch on and off the potential. However, to the best of our knowledge, the stochastic thermodynamics of this class of models has not been considered before. It is interesting to note that the energetic cost of resetting, the energy needed to trap a particle and drag it back to its reset position,   has been measured experimentally in a context where the resetting takes a finite amount of time \cite{friedman2020exp}. The distribution of the work required to reset the
system under a non-instantaneous resetting protocol has also been investigated theoretically, very recently in \cite{Deepak2022_work}.

This paper is organized as follows. In Section \ref{sec:model}, we introduce the model and derive expressions for the entropy production rate per individual reset as a trajectory-wise quantity. In Section \ref{sec:distribution}, we detail how the full distribution of entropy production can be obtained in this system and in Section \ref{sec:exact} we compute this distribution at late times for two specific resetting distributions. We end with a discussion in Section \ref{sec:discussion}. The appendices carry further details of the calculations.

\section{Entropy production of resetting}
\label{sec:model}

\begin{figure}[t]
\centering
\includegraphics[width=9cm]{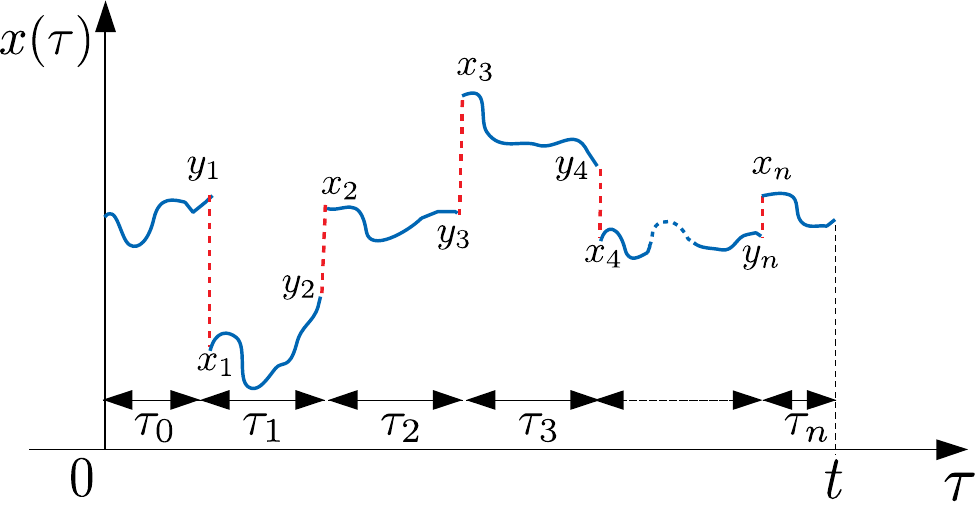}  
\caption{\label{fig:trajectory} Schematic representation of the resetting process with $n$ resetting events (red dashed lines) and total duration $t$. The position of the process immediately before (after) the $i$-th resetting event is $y_i$ ($x_i$). The time between the $i$-th and the $(i+1)$-th resetting is denoted by $\tau_i$. }
\end{figure}

We consider a one-dimensional overdamped Brownian particle in a potential $V(x)$. We assume that the particle undergoes stochastic resetting at a constant rate $r$ and that after the resetting event the new position of the particle is independently drawn from the resetting distribution $p_R(x)$. In other words, in a small time interval $dt$, the position $x(\tau)$ of the particle evolves according to the stochastic rule
\begin{equation}
    x(\tau+dt)=\begin{cases}
    x(\tau)-V'(x)dt+\sqrt{2D}\eta(\tau)dt\quad&\text{with probability }1-rdt\,,\\
    \\
    x_{\rm res}\quad&\text{with probability }rdt\,,\\
    \end{cases}
    \label{eq:imperfect_resetting}
\end{equation}
where $D>0$ is the diffusion constant, $\eta(\tau)$ is Gaussian white noise with zero mean and correlator $\langle \eta(\tau)\eta(\tau')\rangle=\delta(\tau-\tau')$, and $x_{\rm res}$ is a random variable drawn from the probability density function (PDF) $p_R(x)$. The diffusion constant is related to the temperature $T$ of the external bath by the Einstein relation $D=k_B T/\gamma$, where $k_B=1$ is Boltzmann's constant and $\gamma=1$ is the friction coefficient. For a schematic representation of the process, see  Fig.~\ref{fig:trajectory}. Note that the case of resetting to a fixed position is recovered by choosing $p_R(x)=\delta(x-x_0)$. For instance, the resetting dynamics could be implemented by switching on a different potential $U(x)$ and letting the particle reach thermal equilibrium, resulting in the Boltzmann weight $p_R(x)\sim e^{-U(x)/T}$. However, here we do not describe the relaxation of the particle in the resetting potential and assume it to be instantaneous. This approximation corresponds to the limit where the relaxation rate of the particle in the potential $U(x)$ is large compared to the resetting rate $r$.

The evolution of the PDF $p(x,t)$ of the position $x$ of the particle is described by the ''augmented" Fokker-Planck equation (FPE) \cite{evans2011reset,evans2011optimal}
\begin{equation}
\partial_t p(x,t)=-\partial_xj(x,t)-rp(x,t)+rp_R(x)\,,
\label{FPEE}
\end{equation}
where 
\begin{equation}
j(x,t)=- D\partial_x p(x,t)-V'(x)p(x,t)\,
\end{equation}
is the local probability current. For late times, this process reaches the nonequilibrium steady state (see Appendix \ref{app:ss})
\begin{equation}
p_{\rm st}(x)=r\int_{0}^{\infty}d\tau~e^{-r\tau}\int_{-\infty}^{\infty}dx_0~G(x|\tau,x_0)p_R(x_0)\,,
\label{eq:steady_state}
\end{equation}
where $G(x|\tau,x_0)$ is the propagator of Brownian motion in the presence of a potential $V(x)$, i.e., the probability density that the particle goes from $x_0$ to $x$ in a time $\tau$. This steady state is characterized by a non-vanishing local probability current 
\begin{equation}
j_{\rm st}(x)=- D\partial_x p_{\rm st}(x)-V'(x)p_{\rm st}(x)\,.
\end{equation}
As a consequence, the system is out of equilibrium.

The total entropy production rate for steady-state diffusive processes is usually estimated as \cite{Seifert:2012stf}
\begin{equation}
\langle \dot{S}_{\rm Local} \rangle=\frac{1}{D}\int_{-\infty}^{\infty} dx~\frac{j^2_{\rm st}(x)}{p_{\rm st}(x)}\,.
\label{S2}
\end{equation}
Eq.~(\ref{S2}) is motivated for systems satisfying a local Langevin dynamics \cite{qian2001mesoscopic}, where it can be shown to arise from averaging over the entropy associated with each possible microscopic trajectory \cite{Seifert:2005epa}. However, this definition fails to capture the total entropy production whenever the dynamics of the system allows for nonlocal jumps \cite{Busiello2019coarse}, as in Eq.~\eqref{eq:imperfect_resetting}. A more general characterization of the total entropy production which relates it to time-reversal symmetry breaking is the log ratio of the probability $\mathcal{P}_{\rm forward}$ of a given trajectory of duration $t$ to the probability $\mathcal{P}_{\rm backward}$ of its time-reversed counterpart \cite{Seifert:2005epa}
 \begin{equation}
\langle \dot{S}_{\mathrm{Total}} \rangle=  \lim_{t\to \infty}\frac{1}{t} \left\langle  \ln{ \left[\frac{\mathcal{P}_{\rm forward}}{\mathcal{P}_\text{backward}} \right]}\right\rangle.
\label{S3}
\end{equation}
This identification of the entropy as quantifying the irreversibility of the dynamics is a cornerstone of the field of stochastic thermodynamics, but this expression becomes singular for fully irreversible processes, such as perfect resetting, where $\mathcal{P}_\text{backward}=0$.

For systems with local Langevin dynamics, the definitions in Eq.~(\ref{S2}) and (\ref{S3}) are equivalent \cite{Seifert:2005epa}. 
A natural question that can arise is hence, to what extent Eqns.~(\ref{S2}) and (\ref{S3}) give different results in the case of the FPE (\ref{FPEE}). This comparison cannot be made in processes with perfect resetting since, as mentioned above, the probability of the time-reversal of any trajectory containing a reset will be strictly zero. Perfect resetting hence adds an essentially irreversible component \cite{Busiello2020uni}, taking it out of the framework of standard stochastic thermodynamics. We are thus motivated to look at the reset mechanism in Eq.~\eqref{eq:imperfect_resetting}, which leads to a well-defined entropy production.

Using the definition in Eq.~\eqref{S3} and applying a path-integral technique, in Appendix \ref{app:path_integral} we split the total entropy production into two contributions 
\begin{equation}
\langle \dot{S}_{\rm Total} \rangle= \langle \dot{S}_R\ \rangle+\langle \dot{S}_m \rangle,
\label{S1}
\end{equation}
where
\begin{equation}
\langle \dot{S}_R \rangle =r\int_{-\infty}^{\infty} dy\int_{-\infty}^{\infty} dx ~p_{\rm st}(y)p_R(x)\log\left[\frac{p_R(x)}{p_R(y)}\right]\,,
\label{Sr}
\end{equation}
is the rate of entropy production due to the nonlocal resetting dynamics and
\begin{equation}
\langle \dot{S}_m \rangle=-\frac{1}{D}\int_{-\infty}^{\infty} dx~V'(x)j_{\rm st}(x)\,,
\label{Sm}
\end{equation}
is the rate of entropy production associated with periods between resetting events. We could now ask how $\langle \dot{S}_{\rm Total} \rangle $ compares with $ \langle \dot{S}_{\rm Local} \rangle$. In what follows we show that in fact  $\langle \dot{S}_{\rm Total} \rangle \geq \langle \dot{S}_{\rm Local} \rangle$. 

\begin{figure}[t]
\centering
\includegraphics[width=8.5cm]{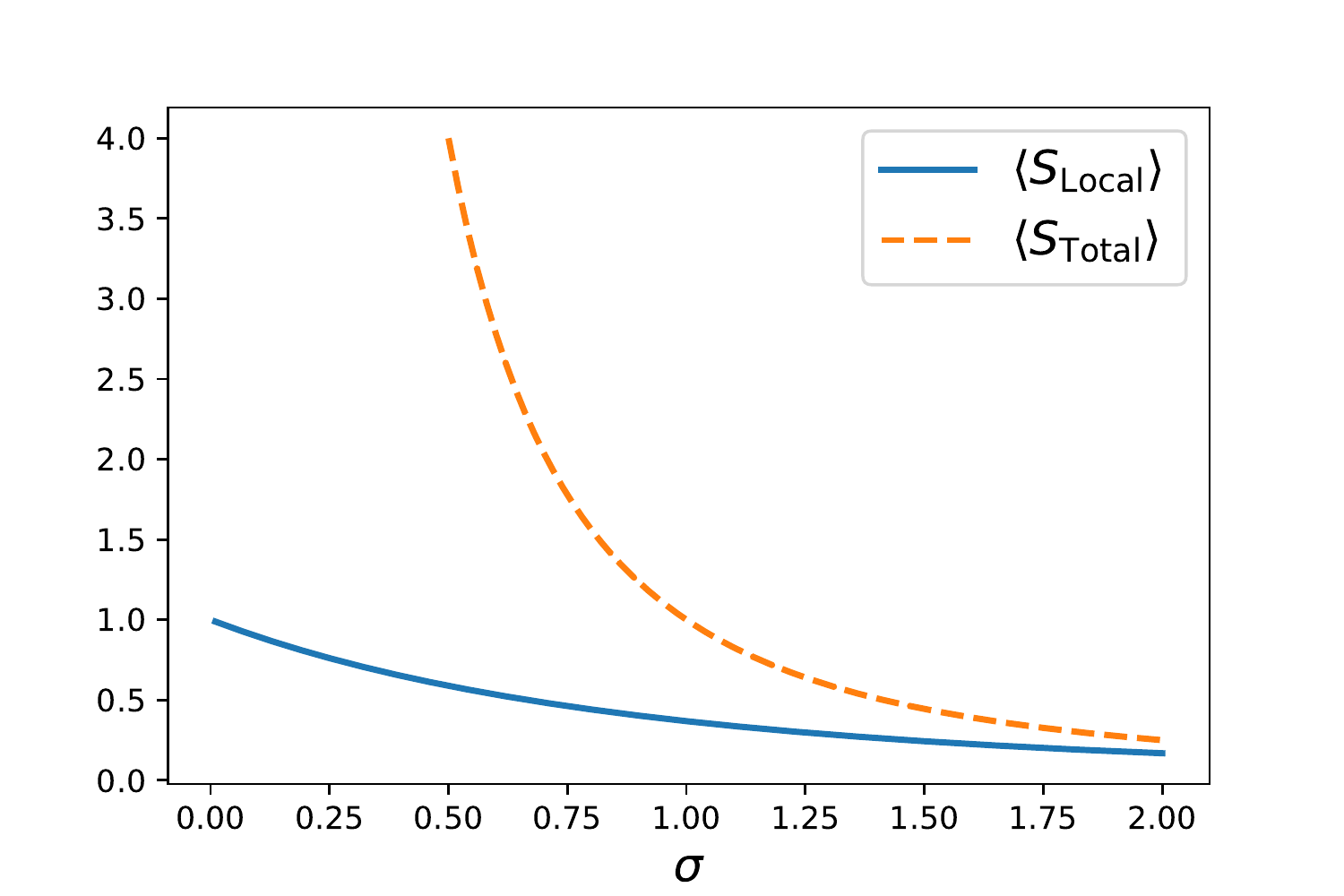}  
\caption{\label{fig:compare} Entropy production as a function of the variance $\sigma$ of the resetting distribution $p_R(x)=e^{-x^2/(2\sigma^2)}/\sqrt{2\pi\sigma^2}$ for a system with $D=r=1$ and $V(x)=0$. The limit $\sigma\to 0$ corresponds to perfect resetting to the origin. In this limit, the local entropy production $\langle S_{\rm Local}\rangle$ (defined in Eq.~\eqref{S2}) converges to a finite value, while the total entropy production $\langle S_{\rm Total}\rangle$ (defined in Eq.~\eqref{S1}) diverges. }
\end{figure}

Following the standard procedure in \cite{broeck2010FP,fuchs2016stochastic}, we consider the time-dependent average system entropy
\begin{equation}
\langle S(t) \rangle=-\int_{-\infty}^{\infty} dx~p(x,t)\log[p(x,t)]\,.
\end{equation}
Differentiating with respect to $t$, we obtain
\begin{equation}
\langle \dot{S}(t) \rangle=-\int_{-\infty}^{\infty} dx~\partial_t p(x,t)\log[p(x,t)]\,.
\end{equation}\label{timeent}
By definition, this system entropy will vanish in the steady state as $t\to \infty$. Using the Fokker-Planck equation Eq.~(\ref{FPEE}) together with the definitions in Eqs.~\eqref{S2} and \eqref{Sm}, we find that in the steady state
\begin{equation}
\langle \dot{S}_{\rm Local} \rangle- \langle \dot{S}_m \rangle +r\int_{-\infty}^{\infty} dx~ p_{\rm st}(x)\log[p_{\rm st}(x)] - r\int_{-\infty}^{\infty} dx~ p_R(x)\log[p_{\rm st}(x)]=0\,.
\end{equation}
Using the definition of the entropy production due to resetting in Eq.~(\ref{Sr}), and writing the  total entropy production as $\langle \dot{S}_{\rm Total} \rangle= \langle \dot{S}_m \rangle+ \langle \dot{S}_R \rangle$, we arrive at
\begin{eqnarray}
\langle \dot{S}_{\rm Total} \rangle- \langle \dot{S}_{\rm Local} \rangle&=&rD_{\rm KL}\left[p_{\rm st}||p_R\right]+rD_{\rm KL}\left[p_R||p_{\rm st}\right]\,,
\label{relation1}
\end{eqnarray}
where
\begin{equation}
D_{\rm KL}\left[p||q\right]=\int_{-\infty}^{\infty} dx~p(x)\log\left[\frac{p(x)}{q(x)}\right]
\end{equation}
is the Kullback–Leibler divergence. Since the right-hand side of Eq.~(\ref{relation1}) is always positive, we find
\begin{equation}
\langle \dot{S}_{\rm Total} \rangle \geq \langle \dot{S}_{\rm Local} \rangle\,.
\label{S1S2}
\end{equation}
Note that when $p_R(x)=\delta(x-x_R)$ the right-hand side in Eq.~\eqref{relation1} diverges while $ \langle \dot{S}_\text{Local} \rangle$ remains finite (see Fig.~\ref{fig:compare}).  Interestingly, the entropy production associated with the nonlocal jumps, given in Eq.~\eqref{relation1}, can be written in terms of the symmetrized Kullback–Leibler divergence between $p_{\rm st}$ and $p_R$, a measure of the distance between the steady state and the resetting distributions.

Eq.~(\ref{S2}) underestimates the total entropy production since it only indirectly incorporates the effect of the nonlocal resetting dynamics from Eq.~(\ref{FPEE}) through the local currents $j_\text{st}(x)$.  The underestimate has exactly the form of the Kullback-Leibler divergence because of the specific form of the nonlocal jump term, that does not depend on the starting position.
Interestingly, a similar reasoning applies also to unidirectional processes, of which resetting is a particular case, where again it is nontrivial to write the rate of change of system entropy in terms of a difference of total and medium entropies. As explained in Appendix \ref{app:stot}, this can
again lead to an underestimate exactly as
in Eq.~(\ref{relation1}).
It is also known that Eq.~(\ref{S2}) can underestimate the total entropy production 
when local or nonlocal
currents are coarse-grained over, indicating the
 loss of information in the coarse-graining process \cite{broeck2010FP,Busiello2019coarse}. Hence, the entropy production of both local and nonlocal currents is correctly accounted for by Eq.~(\ref{S3}) as long as  these are not unidirectional. However, correctly characterizing the total entropy production rate, as well as higher moments of the entropy production, of (seemingly) unidirectional transitions remains an interesting open problem \cite{Busiello2020uni}.

For the model of resetting that we study here, which is not unidirectional, we are able to describe correctly the entropy production rate at the trajectory level. This allows us to compute properties of the full distribution of entropy production and to build an intuition on the mechanisms that lead to dissipation in resetting systems. In particular, we will be able to 
\begin{enumerate}
    \item[i)] characterize the typical fluctuations around the average value of the entropy production,
    \item[ii)] investigate the likelihood of rare fluctuations away from the mean value,
    \item[iii)] identify the features of the trajectories that lead to atypically high or low entropy production.
\end{enumerate}

\section{Distribution of total entropy}
\label{sec:distribution}

As we have seen above, describing the entropy production directly at the trajectory level properly accounts for the instantaneous resetting events. Being a function of a single stochastic realization of the system trajectory, the entropy is a random variable. Here, we present the framework for describing the full statistics of the entropy production, which takes a large-deviation form at late times. In the following section, we  apply this technique to two concrete examples. It is relevant to mention that the large deviations of different observables of processes with perfect resetting have been previously investigated in Refs.~\cite{meylahn2015LD,harris2017phase,coghi2020large,SM22}, where dynamical phase transitions have been observed.

We consider a long time series $x(\tau)$ of total duration $t$, evolving in time according to Eq.~\eqref{eq:imperfect_resetting}. We assume that at the initial time, the process is already in the steady state (this assumption can be lifted when discussing the late-time properties). We want to compute the distribution of the total entropy production $S_{\rm Total}$, defined as 
\begin{equation}
S_{\rm Total}=\log\left(\frac{\mathcal{P}[x(\tau)]}{\mathcal{P}[x(t-\tau)]}\right)\,,
\end{equation}
where $\mathcal{P}[x(\tau)]$ is the probability of observing the $x(\tau)$ (with $0<\tau<t$), while $\mathcal{P}[x(t-\tau)]$ is the probability to observe the time-reversed trajectory $x(t-\tau)$. In Appendix \ref{app:path_integral}, using a path-integral formalism, we show that the total entropy production can be written as the sum of the entropy production $S_R$ due to  resetting, the change of entropy $\Delta S$ of the particle, and the medium entropy production $S_m$:
\begin{equation}
S_{\rm Total}=S_R+\Delta S + S_m\,.
\label{eq:Stot_decomposition}
\end{equation} 
All the terms contributing to the total entropy are random variables that depend on the particular system trajectory. Between resetting events, energy is dissipated by the particle into the bath in the presence of a potential. The corresponding entropy produced is
\begin{equation}
S_m=-\sum_{i=1}^{n+1}\frac{ V(y_{i}) - V(x_{i-1})}{T}\,,
\label{Sm1}
\end{equation}
where $T$ is the temperature of the thermal bath and $n$ is the number of resetting events. Here $y_i$ is the position of the particle right before the $i$-th resetting event, and $x_i$ is the position of the particle right after the $i$-th resetting event (see Fig.~\ref{fig:trajectory}). We use the notation $x_0=x(0)$ and $y_{n+1}=x(t)$. The entropy associated with resetting is governed by the transition probability densities as before, namely
\begin{equation}
S_R=\sum_{i=1}^{n}\log\left(\frac{p_R(x_i)}{p_R(y_i)}\right)\,.
\end{equation}
Moreover, the total change of entropy of the particle is given by
\begin{equation}
\Delta S=-\log[p_{\text{st}}(x(0))]+\log[p_{\text{st}}(x(t))]\,.
\end{equation}
where we used $S = - \log p(x(t),t)$, with $p$ solving the FPE, as the definition of the system entropy.

When the observation time $t$ is large, we expect the system entropy to approach a constant $\Delta S\sim \mathcal{O}(1)$ while the other entropy terms will keep growing as $S_R\sim \mathcal{O}(t)$ and $S_m\sim \mathcal{O}(t)$. Therefore, we can approximate the total entropy as
\begin{equation}
S_{\rm Total}\approx S_R + S_m \approx\sum_{i=1}^{n}\log\left(\frac{p_R(x_i)}{p_R(y_i)}\right) -  \sum_{i=1}^{n}\frac{ V(y_{i}) - V(x_{j-1})}{T} \,.
\label{Stot_1}
\end{equation}
It is useful to rewrite the right-hand side of Eq.~(\ref{Stot_1})  (for $n\geq 2$)
as
\begin{equation} 
S_{\rm Total}\approx \sum_{i=1}^{n-1}\log\left(\frac{p_R(x_i)}{p_R(y_{i+1})}\right) - \sum_{i=1}^{n-1}\frac{ V(y_{i+1}) - V(x_{i})}{T} = \sum_{i=1}^{n-1}s_i\,,
\label{decomposition}
\end{equation}
where we have neglected order-one terms corresponding to the segment of the particle path before the first resetting and after the last resetting, and we have defined the local entropy variables 
\begin{equation}
s_i=\log\left(\frac{p_R(x_i)}{p_R(y_{i+1})}\right) - \frac{ V(y_{i+1}) - V(x_{i})}{T} \,.
\label{eq:local_variables}
\end{equation}
Note that when expressed in this way, the variables $s_1, s_2,\ldots ,s_{n-1}$ correspond to time intervals separated by resetting events. However, these variables are not independent due to the constraint on the total time. In other words, the fact that we are observing a time window of \emph{fixed} duration $t$ constrains the fluctuations of the local entropy variables. Moreover, the number $n$ of such variables, which corresponds to the number of resetting events, is also random.

Let us denote by $\tau_i$ the duration of the time interval between the $i$-th and the $(i+1)$-th resetting event. Since the resetting events occur at a constant rate $r$, we have
\begin{equation}
p(\tau)=r e^{-r\tau}\,.
\end{equation}
Then, the PDF of $s_i$, conditioned on the duration $\tau_i$ of the corresponding interval, reads
\begin{equation}
p(s_i|\tau_i)=\int_{-\infty}^{\infty}dx~p_R(x)\int_{-\infty}^{\infty}dy~G(y|\tau_i,x)~\delta\left(s_i-\log\left(\frac{p_R(x)}{p_R(y)}\right)+ \frac{ V(y) - V(x)}{T}\right),
\label{pstau}
\end{equation}
where $G(y|\tau,x)$ is the propagator of Brownian motion, i.e., the probability density to go from position $x$ to position $y$ in time $\tau$, in the presence of the potential $V(x)$. For instance, in the case of free Brownian motion ($V(x) = 0$) one has \begin{equation}\label{eq:freeprop}
G(x|\tau,y)=\frac{1}{\sqrt{4\pi D\tau}}e^{-(x-y)^2/(4D\tau)}\,.
\end{equation}
The joint PDF of the local entropy variables $s_1\,,s_2\,,\ldots\,,s_{n-1}$ and of the number $n$ of resetting events can be written as
\begin{equation}
P(s_1\,,s_2\,,\ldots \,,s_{n-1}\,,n|t)=\prod_{i=1}^{n-1}\int_{0}^{\infty}d\tau_i~re^{-r\tau_i}p(s_i|\tau_i)\delta\left(\sum_{i=1}^{n}\tau_i-t\right)\,.
\end{equation}
This expression manifestly shows that the fixed-time constraint introduces correlations among the $s_i$ variables. 

We can now write the probability distribution of the total entropy production $S_{\rm Total}$ as
\begin{equation}
P(S_{\rm Total}|t)\approx\sum_{n=2}^{\infty}\prod_{i=1}^{n-1}\int_{0}^{\infty}d\tau_i~\int_{-\infty}^{\infty}ds_i~r e^{-r\tau_i}p(s_i|\tau_i)\delta\left(\sum_{i=1}^{n-1}s_i-S_{\rm Total}\right)
\delta\left(\sum_{i=1}^{n-1}\tau_i-t\right)\,,
\label{PS_deltas}
\end{equation}
where the $\delta$-functions constrains the values taken by the total entropy production $S_{\rm Total}$ and the total time $t$. Note that we have approximated $t=\sum_{i=0}^{n}\tau_i\approx\sum_{i=1}^{n-1}\tau_i$ since $t$ is large.  To decouple the variables $s_i$ and $\tau_i$, we insert the integral representations of the $\delta$-function
\begin{equation}
\delta\left(\sum_{i=1}^{n-1}s_i-S_{\rm Total}\right)=\frac{1}{2\pi i}\int_{\Gamma_1}dq~\exp\left[-q\left(\sum_{i=1}^{n-1}s_i-S_{\rm Total}\right)\right]\,,
\end{equation}
and
\begin{equation}
\delta\left(\sum_{i=1}^{n-1}\tau_i-t\right)=\frac{1}{2\pi i}\int_{\Gamma_2}d\lambda~\exp\left[-\lambda \left(\sum_{i=1}^{n-1}\tau_i-t\right)\right]\,.
\end{equation}
Here $\Gamma_1$ and $\Gamma_2$ are Bromwich contours in the complex plane. Plugging these integral representations into Eq.~\eqref{PS_deltas}, we get
\begin{equation}
P(S_{\rm Total}|t)\approx\frac{1}{2\pi i}\int_{\Gamma_1}dq~e^{qS_{\rm Total}}\frac{1}{2\pi i}\int_{\Gamma_2}d\lambda~e^{\lambda t}\sum_{n=2}^{\infty}\left[r\int_{-\infty}^{\infty}ds~e^{-qs}\int_{0}^{\infty}d\tau~ e^{-(r+\lambda)\tau}p(s|\tau)\right]^{n-1}\,.
\end{equation}
Performing the sum over $n$, we obtain
\begin{equation}
P(S_{\rm Total}|t)\approx \frac{1}{2\pi i}\int_{\Gamma_1}dq~\frac{1}{2\pi i}\int_{\Gamma_2}d\lambda~e^{qS_{\rm Total}+\lambda t}\frac{\tilde{p}(q|r+\lambda)}{1-r\tilde{p}(q|r+\lambda)}\,,
\label{eq:Pstotlam}
\end{equation}
where
\begin{equation}
\tilde{p}(q|\lambda)=\int_{-\infty}^{\infty}ds~e^{-qs}\int_{0}^{\infty}d\tau~e^{-\lambda\tau}p(s|\tau)\,,
\label{pstau_tilde}
\end{equation}
and where $p(s|\tau)$ is given in Eq.~\eqref{pstau}. In the case of free diffusion ($V(x)=0$), using the expression in Eqs.~\eqref{pstau} and \eqref{eq:freeprop} , we find
\begin{equation}
\tilde{p}(q|\lambda)=\int_{-\infty}^{\infty}dx~p_R(x)\int_{-\infty}^{\infty}dy~\frac{1}{2\sqrt{\lambda}}e^{-\sqrt{\lambda}|x-y|}~\exp\left[-q\log\left(\frac{p_R(x)}{p_R(y)}\right)\right]\,.
\label{tildep_def_1}
\end{equation}

Before investigating two specific models, let us mention a general aspect of the distribution of $S_{\rm Total}$. The natural choice in our problem is to consider the total observation time $t$ to be fixed (letting the number $n$ of resetting events fluctuate). In analogy with the literature on run-and-tumble particles \cite{MLDM21}, we will denote this setting as \emph{fixed-$t$ ensemble}. Alternatively, one could consider the (less natural) \emph{fixed-$n$ ensemble}, where exactly $n$ resetting events are observed and the total time $t$ fluctuates.

Since in the fixed-$n$ ensemble the local entropy variables are independent and identically distributed, it is usually easier to perform exact computations. Moreover, several general results are available for the fixed-$n$ ensemble \cite{majumdar2005nature,MLDM21}. The key quantity to investigate in this case is the marginal PDF of the local entropy variable
\begin{equation}
p(s)=\int_{0}^{\infty}d\tau~r e^{-r\tau}p(s|\tau)\,.
\label{ps_integral}
\end{equation}
One important prediction, valid for the fixed-$n$ ensemble, is that, under specific conditions on $p(s)$, the distribution of $S_{\rm Total}$ displays the signatures of a real-space condensation transition \cite{majumdar2005nature}. In other words, above a threshold value $S_{\rm Total}^c$ of the total entropy production, one single local entropy variable $s^*$ will produce a finite fraction $m_c=s^*/S_{\rm Total}$ of the total entropy (with $0<m_c<1$). This critical value $S_{\rm Total}^c$ usually corresponds to rate events in the large-deviation tail of $S_{\rm Total}$. Below the threshold, the different entropy variables $s_1\,,\ldots\,,s_n$ contribute democratically to the total entropy production ($m_c=0$ in the thermodynamic limit $t\to\infty$). The signature of this transition is a singularity in the large-deviation function of the distribution of the total entropy production at $S_{\rm Total}=S_{\rm Total}^c$. Real-space condensation has been observed in a wide range of systems, including mass-transport models \cite{majumdar2005nature}, financial models \cite{filiasi2014concentration}, run-and-tumble particles \cite{GM19,MLDM21,MGM21}, discrete nonlinear Schrödinger equation \cite{gradenigo2021condensation}, among others \cite{SM22}.

In Ref.~\cite{majumdar2005nature}, a general criterion for condensation for a sum of i.i.d. variables was derived in the context of mass-transport models, predicting that condensation occurs if, for large $s$, the distribution $p(s)$ decays slower than any exponential and faster than $1/s^2$, i.e., if
\begin{equation}
e^{-cs}\ll p(s)\ll A/s^2\,,
\label{criterion}
\end{equation}
for any positive constants $c$ and $A$. Even though this result was derived for the fixed-$n$ ensemble, we expect the criterion to be valid for the fixed-$t$ ensemble when the distribution $p(\tau)$ of the resetting times is exponential \cite{MLDM21}. Note that an equivalent criterion for condensation is valid for negative values of $s$.

Another general prediction for fixed-$n$ ensemble is that, in the regime where $S_{\rm Total}\gg \langle S_{\rm Total}\rangle$ (sometimes known as \emph{extreme large-deviation regime}), the total entropy production is dominated by a single local variable \cite{MLDM21}. In other words, when $S_{\rm Total}$ is extremely large, one expects $s^*\approx S_{\rm Total}$ (or equivalently $m_c\approx 1$), where $s^*$ is one of the $n$ local-entropy variables. As a consequence, we expect that (for $S_{\rm Total}\gg \langle S_{\rm Total}\rangle$)
\begin{equation}
    P(S_{\rm Total}|n)\sim p(S_{\rm Total})\,,
    \label{prediction2}
\end{equation}
where $p(s)$ is the local-entropy distribution in Eq.~\eqref{ps_integral}. Note that this regime is approached either with a sharp phase transition (if $p(s)$ satisfies the criterion in Eq.~\eqref{criterion}) or with a smooth crossover. As we will show, this prediction in Eq.~\eqref{prediction2} is not valid in general in the fixed-$t$ ensemble, as a consequence of the constraint on the total time.

\section{Exactly solvable models}
\label{sec:exact}

In this section, we consider two models in which one can exactly compute the distribution of the total entropy production $S_{\rm Total}$ at late times. In both cases, we assume that the particle evolves freely in between resetting events, i.e., that $V(x)=0$. As a consequence, the medium entropy production vanishes between resetting events, i.e., $S_m=0$. For late times, the distribution of the $S_{\rm Total}$ assumes a large-deviation form. Below we compute exactly the corresponding large-deviation function for two different choices of the resetting distribution $p_R(x)$. These large-deviation functions provide information about the likelihood of (both typical and rare) fluctuations around the average entropy production. Moreover, investigating the large deviations of these models, we will also understand which type of trajectories contribute to atypically large values of $S_{\rm Total}$. In this section, we will only present the main results. The details of the computations are given in Appendix \ref{app:computations_ld}.

\subsection{Exponential resetting distribution}

We focus now on the exponential resetting distribution
\begin{equation}
p_R(x)=\frac{a}{2}e^{-a|x|}\,.
\label{PRexp}
\end{equation}
For simplicity, we set $a=D=r=1$. Using Eq.~\eqref{eq:steady_state}, we find that the steady-state distribution of the process  reads
\begin{equation}
p_{\rm st}(x)=\int_{0}^{\infty}d\tau~e^{-\tau}\int_{-\infty}^{\infty} dx_R~\frac{1}{2}e^{-|x_R|}\frac{1}{\sqrt{4\pi  \tau}}e^{-(x-x_R)^2/(4\tau)}=\frac{1}{4}e^{-|x|}(1+|x|)\,.
\end{equation}
The average rate of entropy production due to the resetting process can be computed using Eq.~\eqref{Sr}, yielding
\begin{equation}
\langle\dot{S}_R\rangle=\int dx p_R(x)\log\left[p_R(x)\right]-\int dx p_{\rm st}(x)\log\left[p_R(x)\right]=\frac12\,.
\label{eq:avg_SR_exp}
\end{equation}
Therefore, in the large-$t$ limit, we expect 
\begin{equation}
 \langle S_{\rm Total} \rangle \approx \int_{0}^{t}d\tau~ \langle \dot{S}_R \rangle = \frac{t}{2}\,.
\label{avg_Stot}
\end{equation}
For this model, the PDF of the local entropy production variables can be computed using Eqs.~\eqref{pstau} and \eqref{ps_integral} and reads
\begin{equation}
p(s)=\begin{cases}
    \frac23 e^{-s}\,\quad\text{ for }s>0\,,\\
    \\
    \frac23 e^{2s}\,\quad\text{ for }s<0\,.\\
\end{cases}
\end{equation}
This distribution does not satisfy the criterion in Eq.~\eqref{avg_Stot}, so we do not expect to observe a condensation transition.

\begin{figure}[t]
\begin{center}
\includegraphics[scale=0.7]{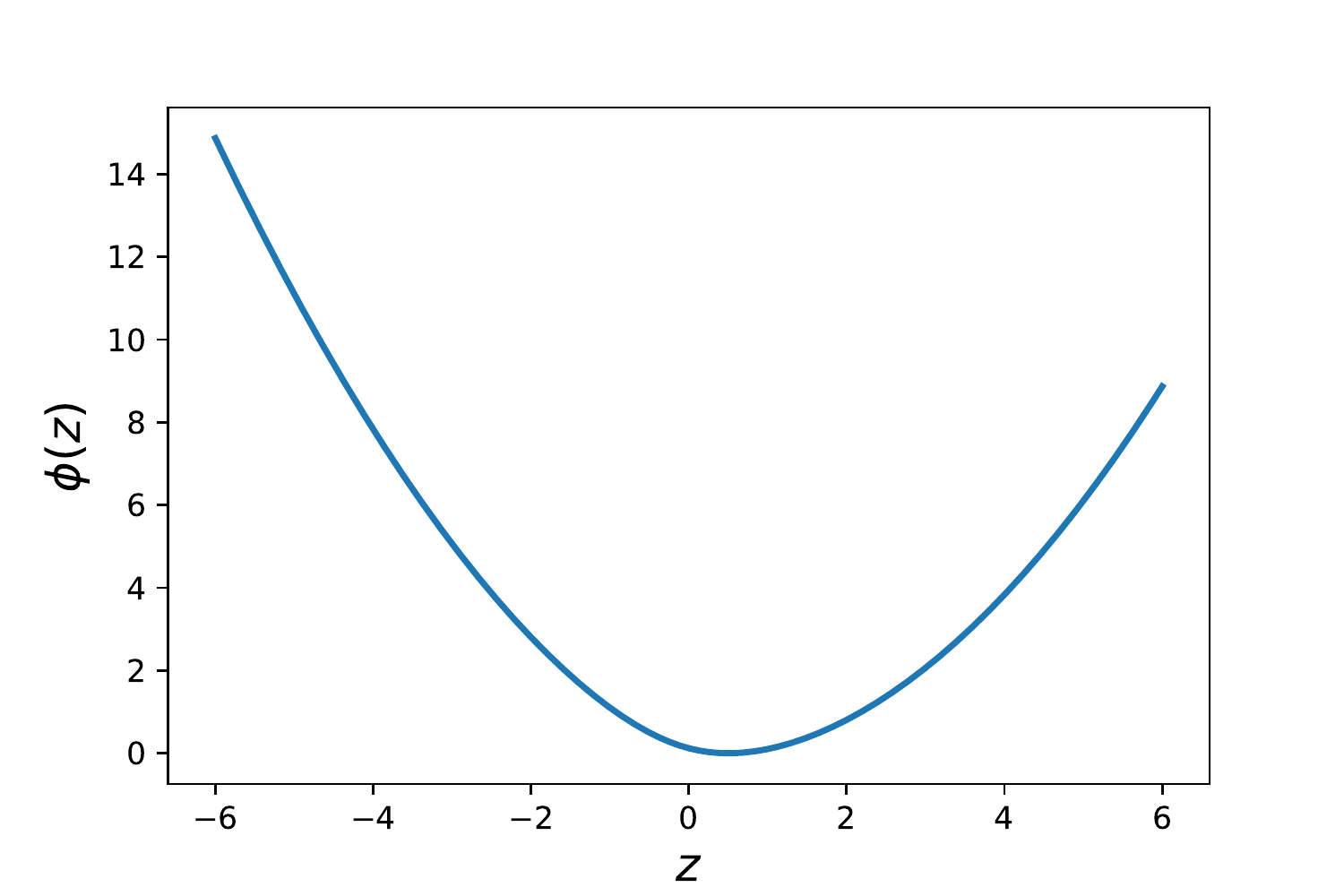}  
\end{center}
\caption{\label{fig:phi} Rate function $\phi(z)\approx-\log(P(S_{\rm Total}|t))/ t$ as a function of $z=S_{\rm Total}/t$. The rate function has a minimum at the typical value $z=1/2$, where $\phi(z)=0$. }
\end{figure}

\begin{figure}[t]
\begin{center}
\includegraphics[scale=0.7]{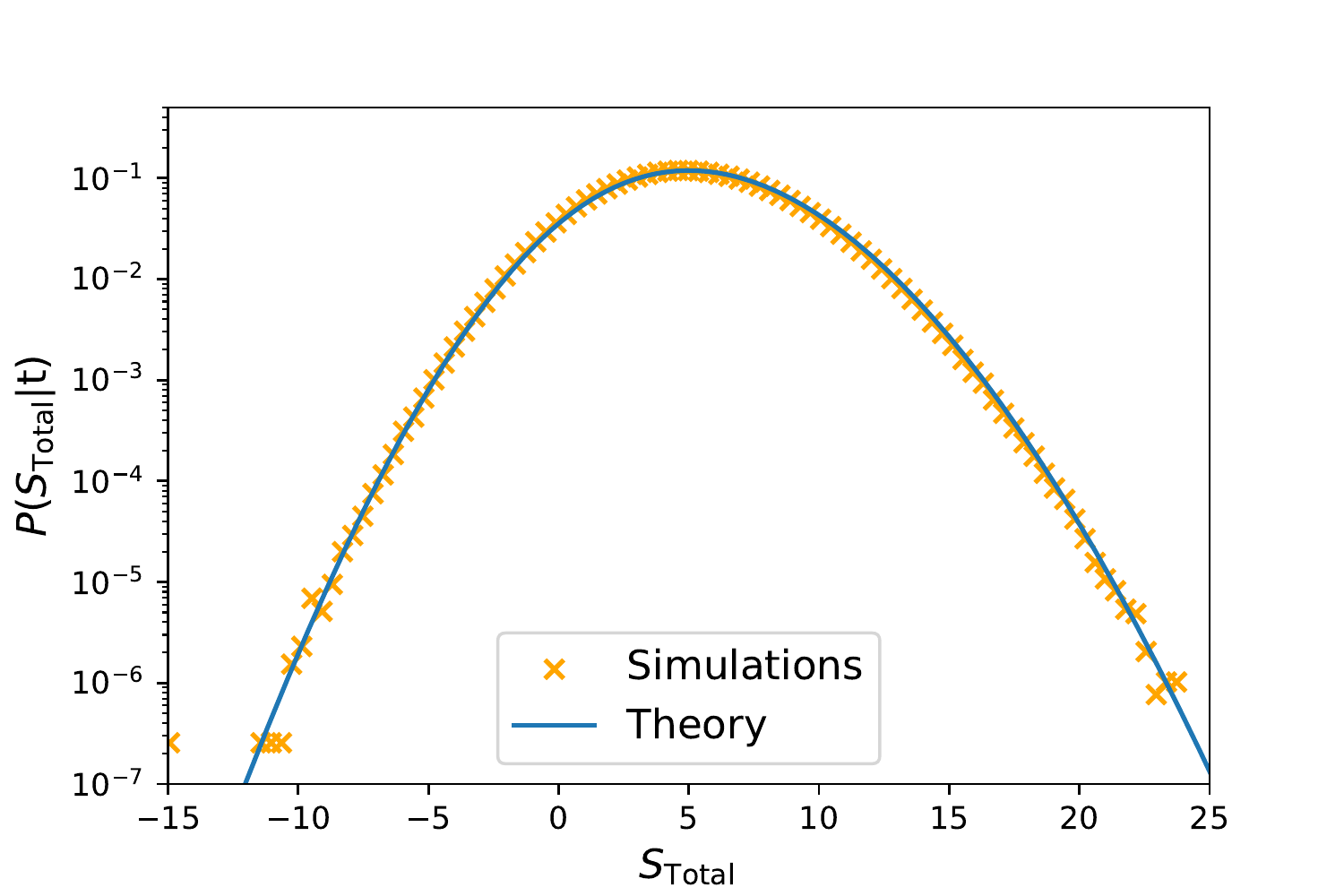}  
\end{center}
\caption{\label{fig:distr_exp} Probability density function $P(S_{\rm Total}|t)$ as a function of the total entropy production $S_{\rm Total}$ for the exponential resetting distribution $p_R(x)=e^{-|x|}/2$. The continuous blue line corresponds to the result in Eq.~\eqref{eq:LDF_exp} (valid for large $t$) while the symbols display the results of numerical simulations with $r=D=1$, $t=10$ and $10^7$ samples.}
\end{figure}

Fluctuations around the average value in Eq.~\eqref{avg_Stot} can be characterized by considering the full distribution of $S_{\rm Total}$. Using the relation in Eq.~\eqref{eq:Pstotlam}, in Appendix \ref{app:computations_ld} we show that the distribution of $S_{\rm Total}$ for late times assumes the large-deviation form (valid for $t\to \infty$, $S_{\rm Total}\to \infty$ with $S_{\rm Total}/t$ fixed)
\begin{equation}
P(S_{\rm Total}|t)\sim \exp\left[-t\phi\left(\frac{S_{\rm Total}}{t}\right)\right]\,,
\label{eq:LDF_exp}
\end{equation}
where the rate function reads
\begin{equation}
    \phi(z)=-\min_{\lambda>\lambda_0}\left[\frac{1}{2}\left[1-\operatorname{sign}(z)\sqrt{1+\frac{4\lambda^2}{1+\lambda-\sqrt{1+\lambda}}}~\right]z+\lambda \right]\,,
\end{equation}
and 
\begin{equation}
\lambda_0=\frac{1}{12}\left[-2-\frac{23 }{(19+12\sqrt{87})^{1/3}}+(19+12\sqrt{87})^{1/3}\right]\approx-0.12097\,.
\end{equation}
This rate function $\phi(z)$ is shown in Fig.~\ref{fig:phi}. We observe that $\phi(z)$ has a unique zero at $z=1/2$, which corresponds to the average value $\langle S_{\rm Total}\rangle\approx t/2$ for large $t$ (see Eq.~\eqref{avg_Stot}). The rate function $\phi(z)$ satisfies the Gallavotti-Cohen theorem \cite{gallavotti1995dynamical}
\begin{equation}
\phi(z)-\phi(-z)=-z\,.
\label{eq:symm1}
\end{equation}
The expression for $P(S_{\rm Total}|t)$ in Eq.~\eqref{eq:LDF_exp} is shown in Fig.~\ref{fig:distr_exp} and is in good agreement with numerical simulations.

The rate function has the following asymptotic behaviors
\begin{equation}
\phi(z)\approx\begin{cases}
\frac{1}{4} z^2-z\,,\quad &\text{ for }z\to -\infty\,\\
\\
\frac{4}{9}\left(z-\frac12\right)^2\,,\quad &\text{ for }z\approx 1/2\\
\\
\frac{1}{4} z^2\,,\quad &\text{ for }z\to \infty\,.
\end{cases}
\label{phi_asymptotics}
\end{equation}
Thus, in the typical regime (close to the typical value $z=1/2$) the distribution of $S_{\rm Total}$ converges for late times to the Gaussian weight
\begin{equation}
P(S_{\rm Total}|t)\sim \exp\left[-\frac{4(S_{\rm Total}-t/2)^2}{9t}\right]\,,
\end{equation}
which is a consequence of the Central Limit Theorem (CLT). From this result, we can also determine the late-time behavior of the variance of $S_{\rm Total}$
\begin{equation}
\langle S_{\rm Total}^2\rangle-\langle S_{\rm Total}\rangle^2\approx \frac{9}{8}t\,.
\label{CLT_11}
\end{equation}

On the other hand, from the large-$z$ behavior of $\phi(z)$ in Eq.~\eqref{phi_asymptotics}, the probability of observing an anomalously large entropy production $S_{\rm Total}\gg \langle S_{\rm Total}\rangle$ decays as
\begin{equation}
P(S_{\rm Total}|t)\sim e^{-S_{\rm Total}^2/(4t)}\,.
\label{extreme_large_dev}
\end{equation}
In other words, the large-$S_{\rm total}$ tail of $P(S_{\rm Total}|t)$ is still Gaussian, but the decay to zero is slower than that predicted by the CLT (see Eq.~\eqref{CLT_11}). Interestingly, one can show that this Gaussian decay corresponds to configurations that are dominated by a single local variable $s^*$, although there is no sharp transition in this case ($s^*$ dominates only asymptotically for $z=S_{\rm Total}/t\to\infty$). Moreover, it is possible to show that the time interval $\tau^*$ associated with $s^*$ approaches asymptotically the total observation time $t$ ($\tau^*\approx t$). In other words, in the extreme large-deviation regime where $S_{\rm Total}\gg t$, the large dissipation is associated with a very long time interval $\tau^*\approx t$ without resetting events. During this interval $\tau^*$, the particle starts from some position $x^*\sim \mathcal{O}(1)$ and propagates ballistically to the final position $y^*\approx S_{\rm Total}\sim \mathcal{O}(t)$, before resetting. Note that this extreme large-deviation tail in Eq.~\eqref{extreme_large_dev} is manifestly different from the one that one would expect in the fixed-$n$ ensemble (see Eq.~\eqref{prediction2}), i.e., when the number $n$ of local entropy variables is fixed. This is a consequence of the constraint on the total time ($\tau^*<t$).

Similarly, the probability of observing configurations in which the entropy production is large and negative decays as (for $|S_{\rm Total}|\gg t$)
\begin{equation}
P(S_{\rm Total}|t)\sim e^{-S_{\rm Total}^2/(4t)+S_{\rm Total}}\,,
\end{equation}
in agreement with the Gallavotti-Cohen theorem. The mechanism for such rare negative events is similar to that of their positive counterparts. One still has a dominant local variable $s^*$, with $\tau^*\approx t$. However, for negative values of $S_{\rm Total}$ one has $x^*\sim \mathcal{O}(t)$ and $y^*\sim \mathcal{O}(1)$.

\subsection{Gaussian resetting distribution}

We next study a Gaussian resetting distribution
\begin{equation}
p_R(x)=\frac{1}{\sqrt{2\pi\sigma^2}}e^{-x^2/(2\sigma^2)}\,.
\label{PRGauss}
\end{equation}
For simplicity, we will work in units where $r = \sigma = D = 1$. The steady-state distribution  can be computed using Eq.~\eqref{eq:steady_state} and reads
\begin{eqnarray}
p_{\rm st}(x)&=&\frac14 e^{1/2- |x|}\left[1+\operatorname{erf}\left(\frac{|x|-1}{\sqrt{2}}\right)+e^{2|x|}\operatorname{erfc}\left(\frac{|x|+1}{\sqrt{2}}\right)\right]\,,
\label{pst_gaussian_2}
\end{eqnarray}
From this, the average rate of entropy production due to the resetting process can be computed directly using Eq.~\eqref{Sr}, yielding
\begin{equation}
\langle \dot S_R \rangle=\int dx~ p_R(x)\log\left[p_R(x)\right]-\int dx ~p_{\rm st}(x)\log\left[p_R(x)\right]=1\,.
\end{equation}
Therefore, in the late-time limit, we find
\begin{equation}
\langle S_{\rm Total}\rangle\approx \int_{0}^{t}d\tau~ \langle \dot S_R \rangle = t\,.
\end{equation}

\begin{figure}[t]
\begin{center}
\includegraphics[scale=0.7]{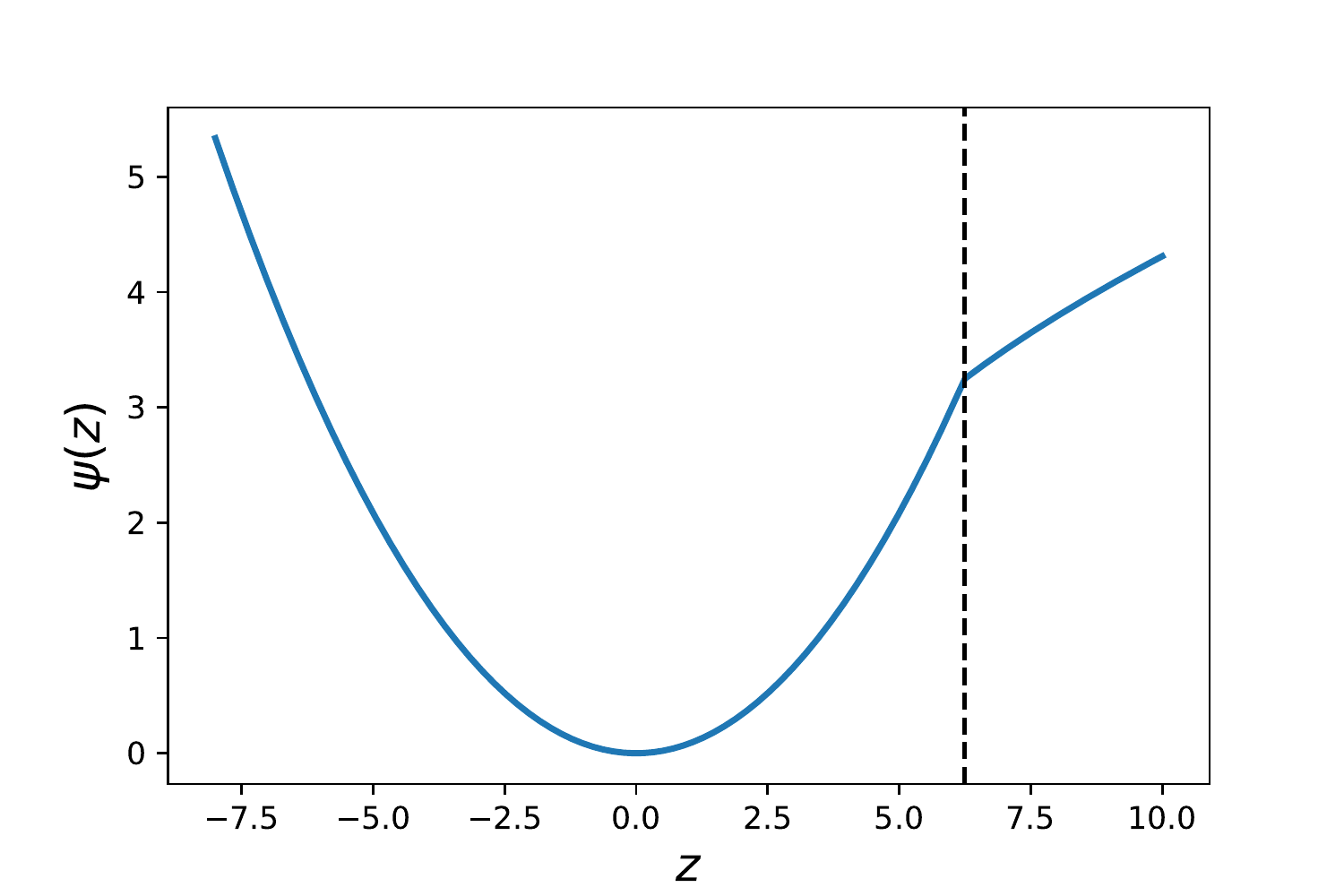}  
\end{center}
\caption{\label{fig:phi2} Rate function $\psi(z)\approx-\log(P(S_{\rm Total}|t))/ t$ as a function of $z=(S_{\rm Total}-t)/t^{2/3}$. The rate function has a first-order singularity at $z=z_c=3^{5/3}$ (vertical dashed line), corresponding to a first-order condensation transition.}
\end{figure}

For this model, the PDF of the local entropy variables reads (see Appendix \ref{app:computations_ld}),
\begin{equation}
p(s)=\int_{-\infty}^{\infty}dx\frac{1}{\sqrt{2\pi}}e^{-x^2/2}\frac{1}{\sqrt{2s+x^2}}\frac12 \left[e^{-|x-\sqrt{x^2+2s}|}+e^{-|x+\sqrt{x^2+2s}|}\right]\,.
\end{equation}
For large $s$, this distribution has the stretched-exponential tail
\begin{equation}
p(s)\approx \sqrt{\frac{e}{2s}}e^{-\sqrt{2s}}\,.
\end{equation}
Interestingly, this distribution satisfies the criterion in Eq.~\eqref{criterion}, and hence we expect to observe a condensation transition in the large deviation regime.

\begin{figure}[t]
\begin{center}
\includegraphics[scale=0.7]{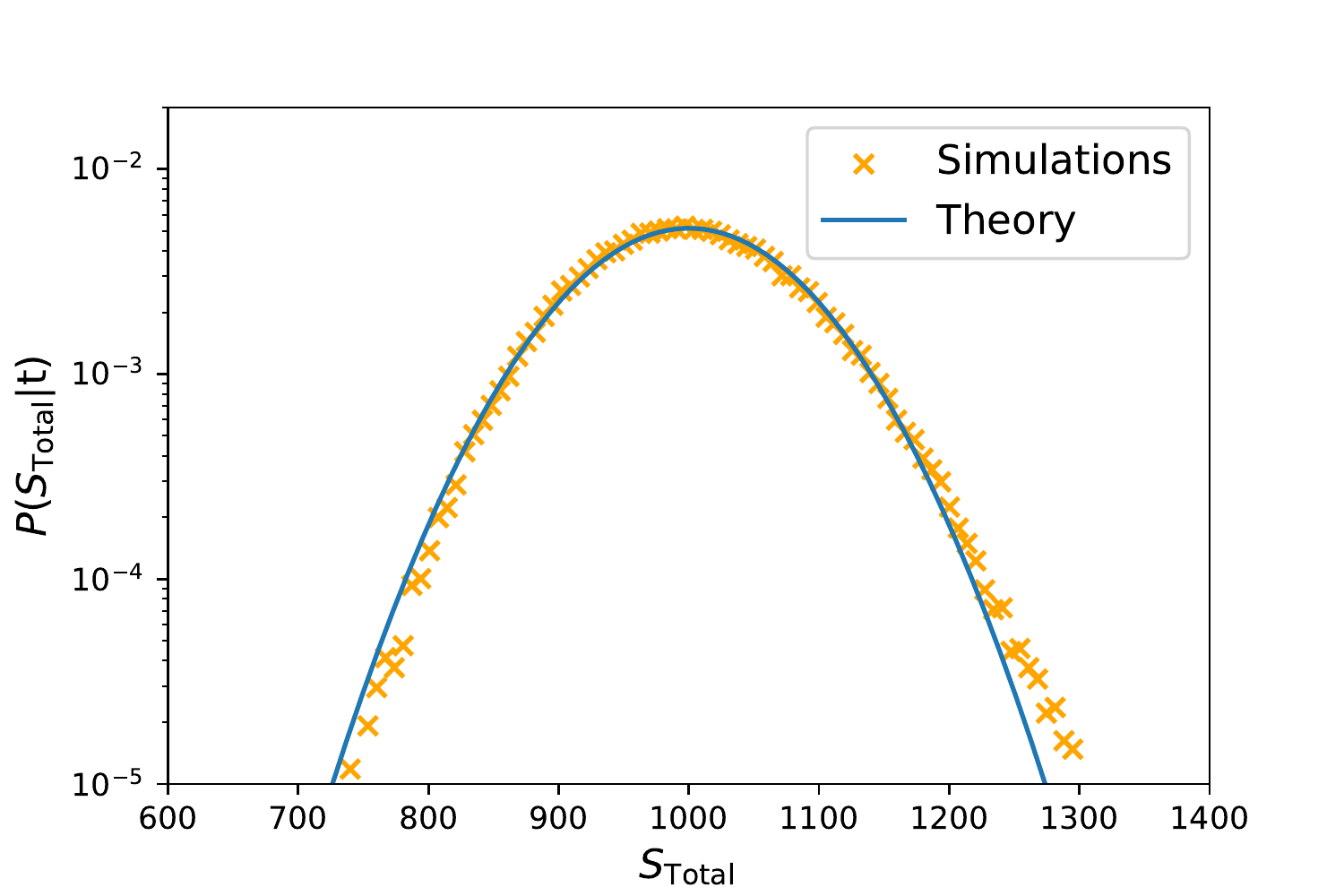}  
\end{center}
\caption{\label{fig:distr_gauss} Probability density function $P(S_{\rm Total}|t)$ as a function of the total entropy production $S_{\rm Total}$ for the Gaussian resetting distribution $p_R(x)=e^{-x^2/2}/\sqrt{2\pi}$. The continuous blue line corresponds to the result in Eq.~\eqref{PST_LDF_1} (valid for large $t$) while the symbols display the results of numerical simulations with $r=D=1$, $t=10^3$ and $10^5$ samples. Deviations from the theoretical curve are finite-size effects.}
\end{figure}

In Appendix \ref{app:computations_ld}, we compute the large deviation function that describes the likelihood of the fluctuations away from the typical value $\langle S_{\rm Total}\rangle\approx t$. In particular, we show that for $|S_{\rm Total}-\langle S_{\rm Total}\rangle|\approx \mathcal{O}(t^{2/3})$
\begin{eqnarray}
 P(S_{\rm Total}|t)\sim \exp\left[-t^{1/3}\psi\left(\frac{S_{\rm Total}-t}{t^{2/3}}\right)\right] \,,
 \label{PST_LDF_1}
\end{eqnarray}
where
\begin{equation}
\psi(z)=\begin{cases}
z^2/12 \quad &\text{ for }z<z_c\,,\\
\\
\chi(z)\quad &\text{ for }z>z_c\,,
\end{cases}
\label{eq:psi1}
\end{equation}
where $z_c=3^{5/3}=6.24025\ldots$ and the function $\chi(z)$ is given in Eq.~\eqref{eq:chi}. This rate function $\psi(z)$ is shown in Fig.~\ref{fig:phi2}. The function $\chi(z)$ has the following asymptotic behaviors (see Appendix \ref{app:computations_ld})
\begin{equation}
\chi(z)\approx
\begin{cases}
    \frac{1}{2~3^{1/3}}(z-z_c)\,\quad & \text{ for }z\to z_c\,,\\
    \\
    \sqrt{2z}-\frac{3}{2z}\,\quad & \text{ for }z\to\infty\,.\\
    \end{cases}
\end{equation}
Thus, at the critical point $z=z_c$, the slope of $\psi(z)$ changes, corresponding to a first-order phase transition. Moreover, for $S_{\rm Total}\to \infty$, we find
\begin{equation}
  P(S_{\rm Total}|t)\sim \exp\left[-\sqrt{2(S_{\rm Total}-t)} \right]\,,
\end{equation}
corresponding to a configuration in which one single local variable is responsible for the total entropy production. In this case, the extreme large-deviation tail ($S_{\rm Total}\gg \langle S_{\rm Total}\rangle$) corresponds to the fixed-$n$ result in Eq.~\eqref{prediction2}. The result for $P(S_{\rm Total}|t)$ in Eq.~\eqref{PST_LDF_1} is shown in Fig.~\ref{fig:distr_gauss}, where it is also compared with numerical simulations finding good agreement.

Interestingly, for this model, a nontrivial large-deviation regime is found for deviations at the scale $S_{\rm Total}-\langle S_{\rm Total}\rangle\sim \mathcal{O}(t^{2/3})$ (instead of the scale $\mathcal{O}(t)$ that is usually observed in large-deviation theory \cite{Touchette:2009lda}). Moreover, we observe that $\psi(z)$ has a first-order singularity at the critical point $z=z_c$, i.e., the first derivative of $\psi(z)$ is discontinuous at $z=z_c$ (see Fig.~\ref{fig:phi2}). This singularity corresponds to a first-order condensation transition: for $z<z_c$ the local entropy variables $s_1\,,\ldots\,,s_n$ contribute to the total entropy by roughly the same amount, while for $z>z_c$ a local entropy variable dominates the total entropy. In other words, in the regime $z>z_c$, where dissipation is atypically large, the most likely configuration corresponds to a single resetting event which contributes macroscopically to the total entropy production. This transition is analogous to the ones described in \cite{majumdar2005nature,GM19,MGM21,SM22}.

As explained in Appendix \ref{app:computations_ld}, this type of transition arises for quantities that can be decomposed into the sum of many independent local variables with a stretched-exponential distribution \cite{nagaev1969integral, brosset2020probabilistic}. In our case, the local variables describe the entropy production associated with an interval between two resetting events. Indeed, from Eq.~\eqref{decomposition}, we have
\begin{equation}
S_{\rm Total}=\sum_{i=1}^{n-1} s_i\,,
\end{equation}
where 
\begin{equation}
s_i=-x_i^2/2+y_i^2/2\,.
\label{s_i}
\end{equation}
 In the condensed phase, for $z>z_c$, one of these local variables becomes anomalously large ($s^*\sim t^{2/3}$) and dominates the total entropy production. From Eq.~\eqref{s_i}, one can show that this occurs in an atypically long period without resetting events ($\tau^*\sim \mathcal{O}(t^{1/3})$) with initial position $x^*\sim \mathcal{O}(1)$ and $y^*\sim \mathcal{O}(t^{1/3})$. However, at variance with the exponential case where $\tau^*\approx t$, in this case the interval $\tau^*$ is small compared to the total time $t$.

Interestingly, for $z<z_c$ we have $\psi(z)=z^2/12$. Using this result in Eq.~\eqref{PST_LDF_1}, we find that for $S_{\rm Total}-t<z_c t^{2/3}$
\begin{equation}
P(S_{\rm Total}|t)\sim \exp\left[-\left(S_{\rm Total}-t\right)^2/(12t)\right]\,.
\end{equation}
In other words, the distribution of $S_{\rm Total}$ remains Gaussian well beyond the regime of validity of the CLT. 
Thus, for late times, the distribution of $S_{\rm Total}$ becomes Gaussian in the typical regime, with variance
\begin{equation}
\langle S_{\rm Total}^2\rangle-\langle S_{\rm Total}\rangle^2\approx 6t\,.
\end{equation}

Note that the large-deviation form in Eq.~\eqref{PST_LDF_1} only describes positive values of $S_{\rm Total}$ (since it characterizes fluctuations around the typical value $t$ of subleading order $t^{2/3}$). To describe configurations where $S_{\rm Total}<0$ one can use the Gallavotti-Cohen theorem \cite{gallavotti1995dynamical}
\begin{eqnarray}
 P(S_{\rm Total}|t)=e^{S_{\rm Total}}P(-S_{\rm Total}|t)\,,
\end{eqnarray}
yielding
\begin{eqnarray}
 P(S_{\rm Total}|t)\sim \exp\left[S_{\rm Total}-t^{1/3}\psi\left(\frac{S_{\rm Total}+t}{t^{2/3}}\right)\right] \,,
 \label{PST_LDF2_main}
\end{eqnarray}
for $S_{\rm Total}<0$. Thus, an analogous phase transition is observed for negative values of the entropy production. The rare configurations with $S_{\rm Total}<-t-z_ct^{2/3}$ are dominated by a single local variable $s^*$, corresponding to a long interval without resetting of duration $\tau^*\sim\mathcal{O}(t^{1/3})$. However, in this case, one finds that the starting position $x^*\sim\mathcal{O}(t^{1/3})$ is atypically large while the final position $y^*\sim \mathcal{O}(1)$.

\section{Discussion}
\label{sec:discussion}

In this paper, we have investigated the thermodynamic cost of resetting with errors. We have provided a framework to compute the full distribution of the entropy production. The proposed framework bypasses the issue of unidirectionality of perfect resetting and allows several thermodynamic quantities to be calculated. In particular, we have considered an overdamped Brownian particle with imperfect resetting. Using a path-integral technique, we have derived exact expressions for the average rate of entropy production. Moreover, we have computed the late-time distribution of the total entropy production, both in the typical and in the large-deviation regimes. We have considered two exactly solvable models corresponding to an exponential and a Gaussian resetting distribution. In the latter case, we have shown that the large-deviation function becomes singular at a critical point, signaling a first-order condensation transition. We have investigated the mechanisms that lead to atypical values of the entropy production for both models.

Many interesting extensions are possible. In particular, while the dynamics between resets was freely diffusive in the present study, one could easily generalize this to more complex dynamics interrupted by resetting. Including multiple timescales or lengthscales could lead to interesting phenomena. Different choices of the resetting distribution could also lead to interesting results.  As we have shown, qualitatively different behaviors can emerge even in the two simple cases studied here. Another relevant direction would be to investigate the optimal resetting protocol that minimizes the dissipation required to perform a given task. This could be done using the optimal-control framework introduced in \cite{de2021resetting}.

Further thermodynamic insights may be obtained by considering the dynamics of the resetting itself. So far, the resetting events have been treated as instantaneous, although imperfect, while realistic implementations would also take a finite amount of time \cite{besga2020PRR,faisant2021_2d,friedman2020exp}. One approach for finite-time resetting explored recently is to implement the reset by switching on a potential and waiting for a first passage to the reset position \cite{Gupta2021_finite}. Another approach is
to model the resetting as ballistic motion towards the initial state \cite{pal2019_finite,maso2019transport, radice2022diffusion}. The effect of non-instantaneous resetting in the context of first-passage problems is also by now well-known \cite{reuveni2016optimal}. The distribution of the work required to reset the system under a non-instantaneous resetting protocol has also been investigated very recently in \cite{Deepak2022_work}. Interestingly, introducing non-instantaneous resetting events and describing the mechanism of resetting would avoid the nonlocal jumps of the dynamics in Eq.~\eqref{eq:imperfect_resetting}.

A framework that combines both imperfect and non-instantaneous resetting is that of intermittent potentials, where a confining potential is switched on and off intermittently to model the resetting events and uninterrupted motion respectively \cite{mercado2020intermittent, mercado2022reducing, santra2021brownian,Deepak2022_work, gupta2020stochastic, gupta2021resetting, alston2022non}. We note that using an intermittent potential as an effective mechanism of resetting is an approach where again the resetting process is no longer strictly unidirectional \cite{alston2022non}. As a consequence, the correct total entropy production also accounts for currents induced by the switching of the potential \cite{alston2022non}. This is hence another scheme to bypass the unidirectionality imposed by perfect resetting and by this means correctly account for the total entropy production. { Indeed, in the case when a potential is switched on for a time long enough for the particle to relax into a Boltzmann state and the relaxation rate is large compared to the resetting rate, predictions for the steady-state mean entropy production based on local currents agree with our estimates \cite{alston2022non}. This shows that including a resetting mechanism in terms of local dynamics will indeed take care of the discrepancy between entropy production based on local currents and that based on time-reversed paths. Furthermore, the inclusion of a resetting distribution not only formally takes care of the problem of absolutely time-irreversible trajectories, but it also enables making physically meaningful predictions.}

Finally, the presented results could be of relevance to experiments. Experimental realization of resetting, for example using optical tweezers, will generically have a finite spread in the resetting position \cite{besga2020PRR,faisant2021_2d}. From knowledge about particle trajectories, in particular the position within an optical resetting trap, the dissipated heat can be estimated and compared with the theoretical predictions. \\

\begin{acknowledgements}
The authors thank S. N. Majumdar, Francesco Coghi and Deepak Gupta for insightful discussions. The authors also acknowledge the Nordita program {\it Are there universal laws in Non-equilibrium Statistical Physics}, May 2022 where this collaboration started.
S.K acknowledges the support of the Swedish Research  Council  through the grant 2021-05070.
K.S.O acknowledges support from the Nordita Fellowship program. Nordita is partially supported by Nordforsk. This work was supported by a Leverhulme Trust International Professorship grant [number LIP-202-014]. For the purpose of Open Access, the authors have applied a CC BY public copyright licence to any Author Accepted Manuscript version arising from this submission.
\end{acknowledgements}

\appendix 

\section{Computation of the steady-state distribution}

\label{app:ss}

In this appendix, we derive an exact formula for the steady-state distribution of Brownian motion in a confining potential $V(x)$ in the presence of constant-rate resetting to the distribution $p_R(x)$. Assuming that the particle is initially located at position $x_i$, one can write the distribution of the particle at time $t$ using the renewal relation
\begin{equation}
    p(x,t)=e^{-rt}G(x|x_i,t)+\int_{0}^{t}d\tau~re^{-r\tau}\int_{-\infty}^{\infty}dx_0~p_R(x_0)G(x|x_0,t)\,,
    \label{eq:renewal_app}
\end{equation}
where $r$ is the resetting rate and $G(x|x_0,t)$ is the probability density that the particle goes from position $x_0$ to position $x$ in time $t$ in the absence of resetting (the precise expression of $G(x|x_0,t)$ will depend on the particular form of the potential $V(x)$). The relation in Eq.~\eqref{eq:renewal_app} can be interpreted as follows. The first term on the right-hand side of \eqref{eq:renewal_app} corresponds to the case where no resetting occurs up to time $t$ (this event occurs with probability $e^{-rt}$). The second term describes the case when at least one resetting event occurs and the last resetting event happens at time $t-\tau$. Taking the limit $t\to \infty$ in Eq.~\eqref{eq:renewal_app}, the first term can be neglected and we obtain the following exact expression for the steady-state distribution
\begin{equation}
    p_{\rm st}(x)=\int_{0}^{\infty}d\tau~re^{-r\tau}\int_{-\infty}^{\infty}dx_0~p_R(x_0)G(x|x_0,t)\,.
\end{equation}

\section{Decomposition of the total entropy production}
\label{app:path_integral}

In this appendix we derive an explicit expression for the total entropy production
\begin{equation}
S_{\rm Total}=\log\left(\frac{\mathcal{P}[x(\tau)]}{\mathcal{P}[x(t-\tau)]}\right)\,.
\label{eq:STOTapp}
\end{equation}
We will use the same notation as in the main text. We denote by $\tau_i$ the duration of the time interval between the $i$-th and the $(i+1)-th$ resetting event (see Fig.~\ref{fig:trajectory}). We also define $t_i=\tau_0+\tau_1+\ldots+\tau_{i-1}$ as the time of the $i$-th resetting event. We denote by $x_i=x(t_i^+)$ ($y_i=y(t_i^-)$) the position of the system right after (before) the $i$-th resetting event. We assume that the system has already reached a steady state at time $\tau=0$ and we observe the trajectory of the system up to time $t$. The probability of observing a trajectory $x(\tau)$ can be written, using the Stratonovich convention, as \cite{Seifert:2012stf}
\begin{equation}
\mathcal{P}[x(\tau)]=\frac{1}{\mathcal{N}}\frac{1}{r}p_{\rm st}(x(0))r e^{-r\tau_0}\prod_{i=1}^n r e^{-r\tau_i}p_R(x(t_i^+))e^{-\mathcal{A}[x(\tau)]}\,,
\label{path_integral}
\end{equation}
where $\mathcal{N}$ is a normalization factor and
\begin{equation}
    \mathcal{A}[x(t)]=\int_{0}^{t}d\tau\left[\frac{1}{4D}\left(\dot{x}+V'(x)\right)^2-\frac12 V''(x)\right]\,,
    \label{action}
\end{equation}
is the action of the trajectory. The last term in the integral in Eq.~\eqref{action} is a consequence of the Stratonovich discretization. Note that the integrand in Eq.~\eqref{action} is discontinuous at the resetting times $t_i$.

Plugging the expression in Eq.~\eqref{path_integral} into the definition in Eq.~\eqref{eq:STOTapp}, we find
\begin{eqnarray}
     \nonumber S_{\rm Total}&=&\left[\log(p_{\rm st}(x(0)))-\log(p_{\rm st}(x(t)))\right]+\sum_{i=1}^n\left[\log(p_R(x(t_i^+)))-\log(p_R(x(t_i^-)))\right]\\&+&\left[\mathcal{A}[x(t-\tau)]-\mathcal{A}[x(\tau)]\right]\,.
    \label{S_TPT}
\end{eqnarray}
The first term on the right-hand side of Eq.~\eqref{S_TPT} is the total change of entropy of the system
\begin{equation}
\Delta S=\log(p_{\rm st}(x(0)))-\log(p_{\rm st}(x(t)))\,.
\end{equation}
The second term is the entropy production due to resetting
\begin{equation}
    S_R=\sum_{i=1}^n\left[\log(p_R(x(t_i^+)))-\log(p_R(x(t_i^-)))\right]=\sum_{i=1}^n\left[\log(p_R(x_i))-\log(p_R(y_i))\right]\,.
    \label{SR_appendix}
\end{equation}
Finally, the last term is the medium entropy production
\begin{equation}
    S_m=\left[\mathcal{A}[x(t-\tau)]-\mathcal{A}[x(\tau)]\right]\,,
\end{equation}
which can be rewritten as
\begin{eqnarray}
     S_m&=&\int_{0}^{t}d\tau\left[\left(\dot{x}(\tau)+V'(x)(\tau)\right)^2/(4D)-V''(x(\tau))/2\right]\\ &-& \nonumber\int_{0}^{t}d\tau\left[\left(\dot{x}(t-\tau)+V'(x(t-\tau))\right)^2/(4D)-V''(x(t-\tau))/2\right]\,.
\end{eqnarray}
Performing the change of variable $\tau\to t-\tau$ in the second interval, we get
\begin{eqnarray}
     S_m&=&\int_{0}^{t}d\tau\left[\left(\dot{x}(\tau)+V'(x)(\tau)\right)^2/(4D)-V''(x(\tau))/2\right]\\ &-& \nonumber\int_{0}^{t}d\tau\left[\left(-\dot{x}(\tau)+V'(x(\tau))\right)^2/(4D)-V''(x(\tau))/2\right]\,.
\end{eqnarray}
Therefore, we find
\begin{eqnarray}
     S_m&=&-\frac{1}{D}\int_{0}^{t}d\tau~\dot{x}(\tau)V'(x(\tau))\,.
     \label{sm_appendix_1}
\end{eqnarray}
Since the integrand is continuous in the intervals $[t_i,t_{i+1}]$, we obtain
\begin{equation}
S_m=-\sum_{i=1}^{n+1}\frac{ V(y_{i}) - V(x_{i-1})}{D}\,,
\end{equation}
where we have used the notation $x_0=x(0)$ and $y_{n+1}=x(t)$.

Finally, let us derive expressions for the average rates of entropy production corresponding to different terms. The average change of the system entropy vanishes in the steady state
\begin{equation}
\langle\Delta S\rangle=\langle\log(p_{\rm st}(x(0)))-\log(p_{\rm st}(x(t)))\rangle=0\,.
\end{equation}
Using Eq.~\eqref{SR_appendix}, the average rate of entropy production due to resetting can be written as 
\begin{eqnarray}
   \nonumber \langle \dot{S}_R\rangle &=&\frac{1}{t}\langle\sum_{i=1}^n\left[\log(p_R(x(t_i^+)-\log(p_R(x(t_i^-)\right]\rangle\\
    &=& \frac{1}{t}\int_{0}^{t}d\tau~\langle\sum_{i=1}^n\delta(\tau-t_i)\left[\log(p_R(x(\tau^+)-\log(p_R(x(\tau^-)\right]\rangle\,.
\end{eqnarray}
In the steady state, the integrand is independent of $\tau$, yielding
\begin{eqnarray}
    \langle \dot{S}_R\rangle=\langle\sum_{i=1}^n\delta(\tau-t_i)\left[\log(p_R(x(\tau^+)))-\log(p_R(x(\tau^-)))\right]\rangle\,,
\end{eqnarray}
where $0<\tau<t$ is some arbitrary instant in time. When $\tau$ coincides with the time of a resetting, which occur at a constant rate $r$, the position $x(\tau^-)$ is distributed according to the steady-state distribution, while $x(\tau^+)$ is drawn from the resetting distribution. Therefore, we obtain
\begin{eqnarray}
    \langle \dot{S}_R\rangle&=&r\int_{-\infty}^{\infty}dx~p_{R}(x)\int_{-\infty}^{\infty}dy~p_{\rm st}(y)\left[\log(p_R(x))-\log(p_R(y))\right]\\
    &=& r\int_{-\infty}^{\infty}dx~p_{R}(x)\log(p_R(x))-r\int_{-\infty}^{\infty}dy~p_{\rm st}(y)\log(p_R(y))\,.
\end{eqnarray}
Using Eq.~\eqref{sm_appendix_1}, the average rate of medium entropy production reads
\begin{eqnarray}
     \langle \dot{S}_m\rangle=-\frac{1}{t}\frac{1}{D}\int_{0}^{t}d\tau~\langle\dot{x}(\tau)V'(x(\tau))\rangle=-\frac1D\langle\dot{x}V'(x)\rangle\,.
\end{eqnarray}
The last average can be evaluated as done in Ref.~\cite{Seifert:2012stf}, yielding
\begin{equation}
\langle \dot{S}_m \rangle=-\frac{1}{D}\int dx~V'(x)j_{\rm st}(x)\,.
\end{equation}

\section{Entropy production for unidirectional transitions}
\label{app:stot}
Perfect resetting is a particular case of a system with unidirectionality, where ,in addition, the transitions are nonlocal. As mentioned in the main text, we will here show how an inequality similar to Eq.~(\ref{S1S2}) can be derived for systems with unidirectional transitions before making connections with the particular case of resetting.
For simplicity, we take the example of a discrete state-space system. Let the system have two different rates $w_{j \rightarrow i}$ and $y_{j \rightarrow i}$ taking it between the same two mesostates $i$ and $j$.  The two different sets of rates can correspond for example to two different physical mechanisms.  { For now, we assume that the $y$-transitions are unidirectional. Namely, $y_{j\to i} \neq 0$ does \emph{not} imply $y_{i\to j} \neq 0$ for all states $\{i,j\}$, as is the case for the bidirectional $w$-transitions.} In the case of the present paper, for example, the $w$-transitions correspond to local diffusive jumps while the $y$-transitions transitions describe perfect resetting to a predefined position. A generic master equation for such a system is
\begin{equation}
\frac{dp_i}{dt} =  \sum_{j=1}^N  \left( w_{j \rightarrow i}p_j - w_{i \rightarrow j}p_i \right) + \sum_{j=1}^N \left( y_{j \rightarrow i}p_j - y_{i \rightarrow j}p_i \right) \,.
\label{ME}
\end{equation}
For such an equation, the  system entropy production rate can be {found by taking the time derivative of the Shannon entropy $\langle S\rangle  =- \sum_{i} p_i \log p_i$ and inserting the above master equation. The result can be}  written as \cite{Busiello2020uni} :
\begin{equation}
\langle \dot{S}(t) \rangle
 =  \sum_{i,j}   w_{j \rightarrow i}p_j \log \frac{ w_{j \rightarrow i}p_j}{ w_{i \rightarrow j}p_i}   -  \sum_{i,j}   w_{j \rightarrow i}p_j \log \frac{ w_{j \rightarrow i}}{ w_{i \rightarrow j}} +
\sum_{i,j}   y_{j \rightarrow i}p_j \log \frac{ p_j}{ p_i} \,.
\label{SME}
\end{equation}
The first term in Eq.~(\ref{SME}) is the total entropy production rate only due to the $w$-transitions and the second is the medium entropy generation rate, again only due to the $w$-transitions. { The last term is due to the unidirectional transitions, which can not be written in the same form as the bidirectional ones due to some transition rates being zero.  It is also worth noting} that the $y$-transitions do play a role in all the terms, via $p$. As argued in \cite{Busiello2020uni}, in steady state, when $ \langle \dot{S}(t) \rangle = 0$, the total entropy production rate of the system can 
be taken to be
\begin{equation}
\langle  \dot S_{\mathrm{Total}}^{\mathrm{uni}}\rangle \equiv \sum_{i,j} w_{j \rightarrow i}{\Tilde{p}}_j \log \frac{ w_{j \rightarrow i} {\Tilde{p}}_j}{ w_{i \rightarrow j}{\Tilde{p}}_i} = \sum_{i,j}   w_{j \rightarrow i}{\Tilde{p}}_j \log \frac{ w_{j \rightarrow i}}{ w_{i \rightarrow j}} -
\sum_{i,j}   y_{j \rightarrow i}{\Tilde{p}}_j \log \frac{ {\Tilde{p}}_j}{{\Tilde{p}}i}  \, ,
\label{ST2}
\end{equation}
where ${\Tilde{p}} $ is the steady-state probability distribution.

This estimate for the entropy production is analogous to the entropy production in Eq.~(\ref{S2}), in the sense that it is the contribution that comes only from the bidirectional transitions \cite{Busiello2020uni}. In fact, if the $w$-transitions describe diffusion, the continuum limit would lead to exactly Eq.~(\ref{S2}). 

However, when the $y$-transitions are not unidirectional, similar to how the resetting transitions are no longer unidirectional after introducing a resetting distribution $p_R(x)$, the standard expressions for entropy production due to Schnakenberg holds \cite{schnakenberg1976network}
\begin{equation}
\langle \dot S_{\mathrm{Total}}^{\mathrm{bi}}\rangle \equiv \sum_{i,j}   w_{j \rightarrow i}p_j \log \frac{w_{j \rightarrow i} p_j}{w_{i \rightarrow j} p_i} + \sum_{i,j}   y_{j \rightarrow i}p_j \log \frac{y_{j \rightarrow i} p_j}{y_{i \rightarrow j} p_i} ,
\label{S_T}
\end{equation}
which is also what one would get from Eq.~(\ref{S3}) by considering individual trajectories. In this case, substituting Eq.~(\ref{S_T}) into Eq.~(\ref{SME}) and taking the steady-state limit, the total entropy production rate in the steady state 
becomes
\begin{equation}
\langle \dot S_{\mathrm{Total}}^{\mathrm{bi}} \rangle = \sum_{i,j}   w_{j \rightarrow i}{\Tilde{p}}_j \log \frac{w_{j \rightarrow i} }{w_{i \rightarrow j} } + \sum_{i,j}   y_{j \rightarrow i}{\Tilde{p}}_j \log \frac{y_{j \rightarrow i} }{y_{i \rightarrow j} } .
\label{ST1}
\end{equation}

 This is analogous to Eq.~(\ref{S1}) in the case of resetting. The case studied in this paper corresponds to the case where the transitions $y_{j\rightarrow i}$ do not depend on the initial state, $y_{j\rightarrow i} = y_{\rightarrow i}$. In this case, the $y$-transitions become nonlocal, in the sense that transitions may occur between two states at arbitrarily large distance in the state space. One can then make the correspondence $y_{\rightarrow i} = rp_R(i)$ and 
${\Tilde{p}}_j = p_{\rm st}(j) $, in which case the two estimates for the total steady-state entropy production rate Eqns. (\ref{ST2}) and (\ref{ST1}) differ exactly by the Kullback-Leibler divergence as in Eq.~(\ref{relation1}). 

The above derivations hence show that a similar problem arises in the more general case of systems with unidirectional transitions, where certain time-reversed paths also have zero probability. By estimating entropy production using only currents that correspond to (statistically) time-reversible dynamics one always finds an underestimate of the entropy production.

\section{Computation of the large deviation function of the entropy production}
\label{app:computations_ld}

In this appendix, we present the computation of the large-deviation function of the entropy production $S_{\rm Total}$ for the two exactly solvable cases presented in the main text.

\subsection{Exponential resetting distribution}

We first consider the exponential resetting distribution 
\begin{equation}
p_R(x)=\frac12 e^{-|x|}\,.
\end{equation}
To compute the distribution of $S_{\rm Total}$ in Eq.~\eqref{eq:Pstotlam} we need to evaluate $\tilde{p}(q,\lambda)$, i.e., the Fourier (with respect to $s$) and Laplace (with respect to $\tau$) transform of the distribution $p(s|\tau)$. Plugging the distribution $p_R(x)$, given in Eq.~\eqref{PRexp}, into Eq.~\eqref{tildep_def_1}, we find
\begin{equation}
\tilde{p}(q|\lambda)=\frac{1}{\lambda-q^2}\left[1-\frac{q}{\sqrt{\lambda}}\frac{1}{1-q+\sqrt{\lambda}}\right]\,.
\end{equation}
Plugging this expression into Eq.~\eqref{eq:Pstotlam}, we obtain
\begin{eqnarray}
\nonumber P(S_{\rm Total}|t)& \approx & \frac{1}{2\pi i}\int_{\Gamma_1}dq~\frac{1}{2\pi i}\int_{\Gamma_2}d\lambda~e^{qS_{\rm Total}+\lambda t}\\ & \times &\frac{\lambda}{\lambda^2+q(1+\lambda-\sqrt{1+\lambda})+q^2(\sqrt{1+\lambda}-\lambda-1)}\,.
\label{eq:Pstotlam2}
\end{eqnarray}

We next compute the integral over $q$. The integral has two poles, at 
\begin{equation}
q_{\pm}(\lambda)=\frac{1}{2}\left[1\pm\sqrt{1+\frac{4\lambda^2}{1+\lambda-\sqrt{1+\lambda}}}\right]\,.
\label{qp_lambda}
\end{equation}
Performing the integral over $q$ and keeping only the exponential part of the integrand, we obtain
\begin{eqnarray}
 P(S_{\rm Total}|t)\sim\frac{1}{2\pi i}\int_{\Gamma}d\lambda~e^{q_-(\lambda) S_{\rm Total}+\lambda t}+\frac{1}{2\pi i}\int_{\Gamma}d\lambda~e^{q_+(\lambda) S_{\rm Total}+\lambda t}\,,
\label{eq:Pstotlam3}
\end{eqnarray}
where $q_-(\lambda)$ and $q_+(\lambda)$ are given in Eq.~\eqref{qp_lambda}. We are interested in the large-deviation regime where $S_{\rm Total}\sim \mathcal{O}(t)$, therefore we introduce the order-one variable $z=S_{\rm Total}/t$ yielding
\begin{eqnarray}
 P(S_{\rm Total}=zt|t)\sim\frac{1}{2\pi i}\int_{\Gamma}d\lambda~e^{t(q_-(\lambda) z+\lambda )}+\frac{1}{2\pi i}\int_{\Gamma}d\lambda~e^{t(q_-(\lambda) z+\lambda )}\,.
\label{eq:Pstotlam4}
\end{eqnarray}
The integral over $\lambda$ can be computed by saddle-point approximation (since $t$ is large). In the case $S_{\rm Total}>0$, it turns out that this expression in Eq.~\eqref{eq:Pstotlam4} is dominated by the unique real saddle point of the second integral, yielding
\begin{eqnarray}
 P(S_{\rm Total}|t)\sim\exp\left[t \min_{\lambda>\lambda_0}\left(q_-(\lambda)z+\lambda\right)\right] \,,
\end{eqnarray}
where the minimization is performed over real values of $\lambda$ and
\begin{equation}
\lambda_0=\frac{1}{12}\left[-2-\frac{23 }{(19+12\sqrt{87})^{1/3}}+(19+12\sqrt{87})^{1/3}\right]=-0.120972\ldots
\end{equation}
The minimization is performed over $\lambda>\lambda_0$ because the integrand in Eq.~\eqref{eq:Pstotlam4} has a branch cut for $\lambda<\lambda_0$ (due to the square root in $q_-(\lambda)$). Thus, the distribution of $S_{\rm Total}$ can be written in the large-deviation form
\begin{eqnarray}
 P(S_{\rm Total}|t)\sim\exp\left[-t \phi(S_{\rm Total}/t)\right]\,,
\label{eq:LD}
\end{eqnarray}
where, for $z>0$,
\begin{equation}
\phi(z)=-\min_{\lambda>\lambda_0}\left(q_-(\lambda)z+\lambda\right)\,.
\label{eq:phiz_pos}
\end{equation}
On the other hand, for $S_{\rm Total}<0$, the integral is dominated by the unique positive saddle point of the first integral, yielding
\begin{equation}
\phi(z)=-\min_{\lambda>\lambda_0}\left(q_+(\lambda)z+\lambda\right)\,,
\label{eq:phiz_neg}
\end{equation}
valid for $z<0$. Thus, the rate function can be written in the compact form
\begin{equation}
    \phi(z)=-\min_{\lambda>\lambda_0}\left[\frac{1}{2}\left[1-\operatorname{sign}(z)\sqrt{1+\frac{4\lambda^2}{1+\lambda-\sqrt{1+\lambda}}}~\right]z+\lambda \right]\,.
\end{equation}

We next compute the asymptotic behaviors of the rate function $\phi(z)$. We first focus on the typical regime where $z\approx 1/2$. It is useful to define the function
\begin{equation}
g_{\pm}(\lambda,z)=q_{\pm}(\lambda)z+\lambda\,.
\label{gpm}
\end{equation}
The typical regime corresponds to small values of $\lambda$, therefore expanding $g_-(\lambda,z)$ in this limit, we find
\begin{equation}
g_-(\lambda,z)\approx (1-2z)\lambda+\frac{9}{2}z\lambda^2\,.
\end{equation}
Minimizing both sides with respect to $\lambda$ and using Eq.~\eqref{eq:phiz_pos}, we obtain
\begin{equation}
\phi(z)\approx \frac{1-4z+4z^2}{18 z}\,,
\end{equation}
which is only valid for $z\approx 1/2$. Finally, setting $z=1/2+\epsilon$ and expanding to leading order for small $\epsilon$, we obtain
\begin{equation}
\phi(z)\approx \frac{4}{9}\epsilon^2=\frac{4}{9}(z-1/2)^2\,.
\end{equation}
Plugging this expression back into the large-deviation form in Eq.~\eqref{eq:LD}, we find
\begin{equation}
P(S_{\rm Total}|t)\sim \exp\left[-\frac{4(S_{\rm Total}-t/2)^2}{9t}\right]\,,
\end{equation}
which is just a consequence of the CLT. From this expression, we can also infer that the variance of $S_{\rm Total}$ grows at late times as
\begin{equation}
\langle S_{\rm Total}^2\rangle-\langle S_{\rm Total}\rangle^2\approx \frac{9}{8}t\,.
\end{equation}

Finally, we also investigate the asymptotic behavior of $\phi(z)$ for large (positive or negative) $z$. Let us consider the case $z>0$. The large-$z$ limit corresponds to $\lambda\to \infty$. Expanding $g_-(\lambda,z)$, defined in Eq.~\eqref{gpm}, for large $\lambda$, we get
\begin{equation}
g_-(\lambda,z)\approx \lambda-\frac{z}{\sqrt{\lambda}}\,.
\end{equation}
Minimizing with respect to $\lambda$ and using the definition of $\phi(z)$ in Eq.~\eqref{eq:phiz_pos}, we find that for large $z$
\begin{equation}
\phi(z)\approx \frac{1}{4} z^2\,.
\end{equation}
The $z\to -\infty$ behavior can be obtaining by using the symmetry in Eq.~\eqref{eq:symm1}, yielding
\begin{equation}
\phi(z)\approx \frac{1}{4} z^2-z\,.
\end{equation}
To summarize, we have shown that the rate function $\phi(z)$ has asymptotic behaviors
\begin{equation}
\phi(z)\approx\begin{cases}
\frac{1}{4} z^2-z\,\quad &\text{ for }z\to -\infty\,\\
\\
\frac{4}{9}\left(z-\frac12\right)^2\,,\quad &\text{ for }z\approx 1/2\\
\\
\frac{1}{4} z^2\,\quad &\text{ for }z\to \infty\,.
\end{cases}
\end{equation}

\subsection{Gaussian resetting distribution}

We next consider the Gaussian resetting distribution
\begin{equation}
p_R(x)=\frac{1}{\sqrt{2\pi }}e^{-x^2/2}\,.
\end{equation}
Let us first investigate the asymptotic behaviors of the distribution $p(s)$ of a local entropy variable, defined as
\begin{eqnarray}
p(s)&=&\int_{0}^{\infty}d\tau~p(\tau)p(s|\tau) \\ \nonumber
&=&\int_0^{\infty}d\tau~r e^{-r\tau}\int_{-\infty}^{\infty}dx\frac{1}{\sqrt{2\pi}}e^{-x^2/2}\int_{-\infty}^{\infty}dy~\frac{1}{\sqrt{4\pi\tau}}e^{-(x-y)^2/(4\tau)}\delta(s-y^2/2+x^2/2)\,.
\end{eqnarray}
where we have used the definition of $p(s|\tau)$ in Eq.~\eqref{pstau}. Performing the integral over $\tau$ and setting $r=1$, we find
\begin{equation}
p(s)=\int_{-\infty}^{\infty}dx\frac{1}{\sqrt{2\pi}}e^{-x^2/2}\int_{-\infty}^{\infty}dy~\frac{1}{2}e^{-|x-y|}\delta(s-y^2/2+x^2/2)\,.
\label{ps_explicit_gauss}
\end{equation}
We first consider the case $s>0$. When $s$ is positive, we can immediately compute the integral over $y$, yielding
\begin{equation}
p(s)=\int_{-\infty}^{\infty}dx\frac{1}{\sqrt{2\pi}}e^{-x^2/2}\frac{1}{\sqrt{2s+x^2}}\frac12 \left[e^{-|x-\sqrt{x^2+2s}|}+e^{-|x+\sqrt{x^2+2s}|}\right]\,.
\end{equation}
Expanding the integrand to leading order for large $s$, we find
\begin{equation}
   p(s)\approx\frac{1}{\sqrt{2s}}\int_{-\infty}^{\infty}dx\frac{1}{\sqrt{2\pi}}e^{-x^2/2}\frac12 \left[e^{-\sqrt{2s}+x}+e^{-x-\sqrt{2s}}\right]=\sqrt{\frac{e}{2s}}e^{-\sqrt{2s}}\,. 
   \label{stretched_exp}
\end{equation}
Therefore, for large $s$, the distribution of the local entropy variable $s$ has a stretched exponential tail. Remarkably, since the total entropy production $S_{\rm Total}$ can be written as a sum of (a random number of) stretched exponential random variables, we expect a condensation transition to occur in the large deviation regime $S_{\rm Total}$, as observed in \cite{GM19} (see also the criterion for condensation in \cite{MLDM21}).

We now consider the case $s<0$. Performing the change of variable $y\to z=y-x$, we rewrite Eq.~\eqref{ps_explicit_gauss} as
\begin{eqnarray}
 p(s)&=&\int_{-\infty}^{\infty}dx\frac{1}{\sqrt{2\pi}}e^{-x^2/2}\int_{-\infty}^{\infty}dz~\frac{1}{2}e^{-|z|}\delta(s-z(z+2x)/2)=\frac{1}{\sqrt{2\pi}}\int_{-\infty}^{\infty}dz~\frac{e^{-|z|}}{2|z|}e^{-(s/z-z/2)^2/2} \nonumber\\
&=& e^{s/2} \frac{1}{\sqrt{2\pi}}\int_{-\infty}^{\infty}dz~\frac{e^{-|z|}}{2|z|}e^{-s^2/(2z^2)-z^2/8}\,.
\end{eqnarray}
From the last expression, we immediately obtain the relation
\begin{equation}
p(s)=e^{s}p(-s)\,,
\end{equation}
in agreement with the Gallavotti-Cohen theorem \cite{gallavotti1995dynamical}. Finally, using Eq.~\eqref{stretched_exp}, we find that for $s\to-\infty$
\begin{equation}
    p(s)\approx \sqrt{\frac{e}{2|s|}}e^{s-\sqrt{2|s|}}\,.
\end{equation}
We conclude that for large negative values of $s$, the PDF $p(s)$ decays exponentially fast.

To compute the distribution of $S_{\rm Total}$, we first need to compute the expression for $\tilde{p}(q|\lambda)$ for this choice of $p_R(x)$.
Plugging the distribution $p_R(x)$, given in Eq.~\eqref{PRGauss}, into Eq.~\eqref{tildep_def_1}, we find
\begin{equation}
\tilde{p}(q|\lambda)=\sqrt{\frac{\pi}{2}}e^{\lambda/(2q(1-q))}\frac{1}{\sqrt{\lambda q(1-q)}}\operatorname{erfc}\left[\sqrt{\frac{\lambda}{2q(1-q)}}\right]\,,
\label{tilde_ps_gauss}
\end{equation}
valid for $0<q<1$. It is also relevant to consider the Fourier transform of the distribution $p(s)$. This quantity is related to $p(s|\tau)$ by
\begin{equation}
p(s)=\int_{0}^{\infty}d\tau~r e^{-r \tau}p(s|\tau)\,.
\end{equation}
Taking a Fourier transform on both sides with respect to $s$, we find
\begin{equation}
\tilde{p}(q)\equiv \int_{-\infty}^{\infty}ds~e^{-qs}p(s)=\int_{-\infty}^{\infty}ds~e^{-qs}\int_{0}^{\infty}d\tau~r e^{-r \tau}p(s|\tau)=r \tilde{p}(q|r)\,.
\end{equation}
Using the expression for $\tilde{p}(q|r)$ in Eq.~\eqref{tilde_ps_gauss} and setting $r=1$, we find
\begin{equation}
\tilde{p}(q)=\int_{-\infty}^{\infty}ds~e^{-qs}p(s)=\sqrt{\pi}e^{1/(2q(1-q))}\frac{1}{\sqrt{ 2q(1-q)}}\operatorname{erfc}\left[\sqrt{\frac{1}{2q(1-q)}}\right]\,.
\label{eq:ptilq}
\end{equation}
Interestingly, this Fourier transform in Eq.~\eqref{eq:ptilq} coincides with the Fourier transform of the jump distribution of the run-and-tumble particle model considered in \cite{GM19}. The equivalence between these two models is quite nontrivial and unexpected. Note that the main difference with the computation done in \cite{GM19} is that here we are fixing the total time, while the number $n$ of local entropy variables can fluctuate. On the other hand, in \cite{GM19}, the number of local variables (the jumps in the language of Refs.~\cite{GM19,MGM21}) is fixed. The two ensembles are usually called ``fixed-$t$'' and ``fixed-$n$'' respectively in the literature of run-and-tumble particles. A condensation transition in the fixed-$t$ ensemble has been observed in Refs.~\cite{MLDM21} (second or higher order transition) and \cite{SM22} (first-order transition).

Plugging the expression for $\tilde{p}(q|\lambda)$, given in Eq.~\eqref{tilde_ps_gauss}, into Eq.~\eqref{eq:Pstotlam}, we obtain
\begin{eqnarray}
\nonumber P(S_{\rm Total}|t)& \approx & \frac{1}{2\pi i}\int_{\Gamma_1}dq~\frac{1}{2\pi i}\int_{\Gamma_2}d\lambda~e^{qS_{\rm Total}+\lambda t}\\ & \times &\frac{\sqrt{\frac{\pi}{2}}e^{(\lambda+1)/(2q(1-q))}\frac{1}{\sqrt{(\lambda+1) q(1-q)}}\operatorname{erfc}\left[\sqrt{\frac{(\lambda+1)}{2q(1-q)}}\right]}{1-\sqrt{\frac{\pi}{2}}e^{(\lambda+1)/(2q(1-q))}\frac{1}{\sqrt{(\lambda+1) q(1-q)}}\operatorname{erfc}\left[\sqrt{\frac{\lambda+1}{2q(1-q)}}\right]}\,.
\label{ltpqlambda1}
\end{eqnarray}

For late times, we expect the distribution of $S_{\rm Total}$ to be peaked around the average value $t$, with Gaussian fluctuations around this value, as a consequence of the CLT. We are interested in computing the large-deviation regime outside of the range of validity of the CLT. We will consider the two cases $S_{\rm Total}>0$ and $S_{\rm Total}<0$ separately. We anticipate that the first case corresponds to the limit $q\to 0$ in Eq.~\eqref{ltpqlambda1}, while the second case corresponds to the limit $q\to 1$. In both cases, we will only consider the late-time regime, corresponding to $\lambda\to 0$.

We start by investigating the first case, corresponding to rare events with the entropy production much larger than the typical value. Expanding the integrand for $q\to 0$, we find
\begin{eqnarray}
 P(S_{\rm Total}|t)\approx \frac{1}{2\pi i}\int_{\Gamma_1}dq~\frac{1}{2\pi i}\int_{\Gamma_2}d\lambda~e^{qS_{\rm Total}+\lambda t}&\frac{\sqrt{\frac{\pi}{2}}e^{(\lambda+1)/(2q)}\frac{1}{\sqrt{(\lambda+1) q}}\operatorname{erfc}\left[\sqrt{\frac{(\lambda+1)}{2q}}\right]}{1-\sqrt{\frac{\pi}{2}}e^{(\lambda+1)/(2q)}\frac{1}{\sqrt{(\lambda+1) q}}\operatorname{erfc}\left[\sqrt{\frac{\lambda+1}{2q}}\right]}\,.
\end{eqnarray}
Performing the change of variable $q\to q'=q/(\lambda+1)$, we obtain
\begin{eqnarray}
 P(S_{\rm Total}|t)\sim \frac{1}{2\pi i}\int_{\Gamma_1}dq'~\frac{1}{2\pi i}\int_{\Gamma_2}d\lambda~e^{q' S_{\rm Total}+\lambda t}&\frac{f(q')}{\lambda+1-f(q')}\,,
\end{eqnarray}
where we have expanded the right-hand side for small $\lambda$ and we have defined
\begin{equation}
f(q)=\sqrt{\frac{\pi}{2}}e^{1/(2q)}\frac{1}{\sqrt{q}}\operatorname{erfc}\left[\sqrt{\frac{1}{2q}}~\right]\,.
\end{equation}
Computing the integral over $\lambda$, we find
\begin{eqnarray}
 P(S_{\rm Total}|t)\sim \frac{1}{2\pi i}\int_{\Gamma_1}dq~f(q)e^{q S_{\rm Total}- t+f(q)t}\,.
 \label{PST_fq_1}
\end{eqnarray}
It is useful to introduce the integral representation of $f(q)$ (which can be verified using Mathematica)
\begin{equation}
f(q)=\int_{0}^{\infty}d\tau~e^{-\tau-q\tau^2/2}\,.
\label{fq_integral_repr}
\end{equation}
Moreover, we also introduce the small-$q$ expansion (valid for $q>0$) 
\begin{equation}
f(q)\approx 1-q+3q^2\,.
\label{fq_expansion}
\end{equation}
Using Eqs.~\eqref{fq_integral_repr} and \eqref{fq_expansion}, the expression in Eq.~\eqref{PST_fq_1} can be rewritten as
\begin{eqnarray}
\nonumber P(S_{\rm Total}|t)&\sim & \frac{1}{2\pi i}\int_{\Gamma_1}dq~\int_{0}^{\infty}d\tau~e^{-\tau-q\tau^2/2} e^{q S_{\rm Total}- t+(1-q+3q^2)t}\\
 &=&\frac{1}{2\pi i}\int_{\Gamma_1}dq~\int_{0}^{\infty}d\tau~e^{-\tau-q\tau^2/2} e^{q S_{\rm Total}+(-q+3q^2)t}\,.
 \label{PST_fq_2}
\end{eqnarray}
Since we are investigating positive large deviations above the average values, we introduce the variable $z$, defined as
\begin{equation}
S_{\rm Total}=t+zt^{\alpha}\,,
\end{equation}
where we recall that $t$ is the average value of $S_{\rm Total}$ and $\alpha$ is a positive constant to be determined. This constant $\alpha$ determines the scale at which the large deviations are observed. Thus, Eq.~\eqref{PST_fq_2} becomes
\begin{eqnarray}
 P(S_{\rm Total}|t)&\sim\frac{1}{2\pi i}\int_{\Gamma_1}dq~\int_{0}^{\infty}d\tau~e^{-\tau-q\tau^2/2+q zt^{\alpha}+3q^2 t}\,.
 \label{PST_fq_3}
\end{eqnarray}
Performing the integral over $q$, we find
\begin{eqnarray}
 P(S_{\rm Total}|t)&\sim \int_{0}^{\infty}d\tau~e^{-\tau-(\tau^2-2t^\alpha z)^2/(48t)}\,.
 \label{PST_fq_4}
\end{eqnarray}
It is useful to perform the change of variables $\tau\to \tau'=\tau t^{-\beta}$, where $\beta>0$ is a constant to be determined, yielding
\begin{eqnarray}
 P(S_{\rm Total}|t)&\sim \int_{0}^{\infty}d\tau'~e^{-\tau' t^{\beta}-({\tau'}^2t^{2\beta}-2t^\alpha z)^2/(48t)}\,.
 \label{PST_fq_5}
\end{eqnarray}
To determine $\alpha$ and $\beta$, we impose that all the terms in the exponent must grow with the same power in $t$, yielding the conditions $\alpha=2\beta$ and $2\alpha-1=\beta$. Solving these equations, we obtain $\alpha=2/3$ and $\beta=1/3$. Thus, Eq.~\eqref{PST_fq_5} reads\begin{eqnarray}
 P(S_{\rm Total}|t)\sim \int_{0}^{\infty}d\tau~e^{-t^{1/3}[\tau +({\tau}^2-2 z)^2/(48)]}\,.
\end{eqnarray}
Finally, performing the integral by saddle-point approximation, we get
\begin{eqnarray}
 P(S_{\rm Total}|t)\sim \exp\left[-t^{1/3}\psi\left(\frac{S_{\rm Total}-t}{t^{2/3}}\right)\right] \,,
 \label{PST_LDF}
\end{eqnarray}
where
\begin{equation}
\psi(z)=\min_{\tau\geq 0}\left[\tau+\frac{1}{48}\left(\tau^2-2z\right)^2\right]\,.
\label{psi_def_1}
\end{equation}
It turns out that for $z<z_c$, where $z_c=3^{5/3}=6.24025\ldots$, the expression in Eq.~\eqref{psi_def_1} is minimal at $\tau=0$, corresponding to 
\begin{equation}
 \psi(z)=z^2/12    
\end{equation}
On the other hand, for $z>z_c$ this expression is minimized by some $\tau^*>0$.
Thus, the rate function $\psi(z)$ can be rewritten as
\begin{equation}
\psi(z)=\begin{cases}
z^2/12 \quad &\text{ for }z<z_c\,,\\
\\
\chi(z)\quad &\text{ for }z>z_c\,,
\end{cases}
\end{equation}
where $z_c=3^{5/3}=6.24025\ldots$. The function $\chi(z)$ can be computed exactly by minimizing analytically the expression in \eqref{psi_def_1}, yielding
\begin{equation}
\chi(z)=\tau^*+\frac{1}{48}((\tau^*)^2-2z)^2\,,
\label{eq:chi}
\end{equation}
where $\tau^*$ is a function of $z$ and reads
\begin{equation}
\tau^*(z)=\frac{6^{1/3}a(z)^2+6^{2/3}z}{3a}\,,
\end{equation}
where
\begin{equation}
a(z)=\left(\sqrt{3}\sqrt{243-2z^3}-27\right)^{1/3}\,.
\end{equation}

We next compute the asymptotic behaviors of $\chi(z)$. The variable $\tau^*(z)$ satisfies the equation
\begin{equation}
\tau^3-2z\tau+12=0\,.
\end{equation}
Solving this equation for large $z$, we find
\begin{equation}
\tau^*(z)\approx \sqrt{2z}-\frac{3}{z}\,.
\end{equation}
Plugging this approximation into the expression for $\chi(z)$ in Eq.~\eqref{eq:chi}, we find 
\begin{equation}
    \chi(z)\approx \sqrt{2z}-\frac{3}{2z}\,.
\end{equation}
By expanding $\chi(z)$ close to the critical point $z_c$, we find
\begin{equation}
    \chi(z)\approx \frac{1}{2~3^{1/3}}(z-z_c)
\end{equation}

To investigate the large-deviation behavior $S_{\rm Total}< 0$, we perform the change of variable $q\to (1-q)$ in Eq.~\eqref{ltpqlambda1}, yielding
\begin{eqnarray}
\nonumber P(S_{\rm Total}|t)& \approx & e^{S_{\rm Total}}\frac{1}{2\pi i}\int_{\Gamma_1}dq~\frac{1}{2\pi i}\int_{\Gamma_2}d\lambda~e^{q(-S_{\rm Total})+\lambda t}\\ & \times &\frac{\sqrt{\frac{\pi}{2}}e^{(\lambda+1)/(2q(1-q))}\frac{1}{\sqrt{(\lambda+1) q(1-q)}}\operatorname{erfc}\left[\sqrt{\frac{(\lambda+1)}{2q(1-q)}}\right]}{1-\sqrt{\frac{\pi}{2}}e^{(\lambda+1)/(2q(1-q))}\frac{1}{\sqrt{(\lambda+1) q(1-q)}}\operatorname{erfc}\left[\sqrt{\frac{\lambda+1}{2q(1-q)}}\right]}=e^{S_{\rm Total}}P(-S_{\rm Total}|t)\,.
\end{eqnarray}
This relation is the Gallavotti-Cohen theorem \cite{gallavotti1995dynamical}. Using this result, we find that for $S_{\rm Total}<0$
\begin{eqnarray}
 P(S_{\rm Total}|t)\sim \exp\left[S_{\rm Total}-t^{1/3}\psi\left(\frac{t-S_{\rm Total}}{t^{2/3}}\right)\right] \,.
\end{eqnarray}

\bibliographystyle{apsrev4-2}
\bibliography{thermodyn_res.bib}

\begin{thebibliography}{58}%
\makeatletter
\providecommand \@ifxundefined [1]{%
 \@ifx{#1\undefined}
}%
\providecommand \@ifnum [1]{%
 \ifnum #1\expandafter \@firstoftwo
 \else \expandafter \@secondoftwo
 \fi
}%
\providecommand \@ifx [1]{%
 \ifx #1\expandafter \@firstoftwo
 \else \expandafter \@secondoftwo
 \fi
}%
\providecommand \natexlab [1]{#1}%
\providecommand \enquote  [1]{``#1''}%
\providecommand \bibnamefont  [1]{#1}%
\providecommand \bibfnamefont [1]{#1}%
\providecommand \citenamefont [1]{#1}%
\providecommand \href@noop [0]{\@secondoftwo}%
\providecommand \href [0]{\begingroup \@sanitize@url \@href}%
\providecommand \@href[1]{\@@startlink{#1}\@@href}%
\providecommand \@@href[1]{\endgroup#1\@@endlink}%
\providecommand \@sanitize@url [0]{\catcode `\\12\catcode `\$12\catcode
  `\&12\catcode `\#12\catcode `\^12\catcode `\_12\catcode `\%12\relax}%
\providecommand \@@startlink[1]{}%
\providecommand \@@endlink[0]{}%
\providecommand \url  [0]{\begingroup\@sanitize@url \@url }%
\providecommand \@url [1]{\endgroup\@href {#1}{\urlprefix }}%
\providecommand \urlprefix  [0]{URL }%
\providecommand \Eprint [0]{\href }%
\providecommand \doibase [0]{https://doi.org/}%
\providecommand \selectlanguage [0]{\@gobble}%
\providecommand \bibinfo  [0]{\@secondoftwo}%
\providecommand \bibfield  [0]{\@secondoftwo}%
\providecommand \translation [1]{[#1]}%
\providecommand \BibitemOpen [0]{}%
\providecommand \bibitemStop [0]{}%
\providecommand \bibitemNoStop [0]{.\EOS\space}%
\providecommand \EOS [0]{\spacefactor3000\relax}%
\providecommand \BibitemShut  [1]{\csname bibitem#1\endcsname}%
\let\auto@bib@innerbib\@empty
\bibitem [{\citenamefont {Evans}\ and\ \citenamefont
  {Majumdar}(2011{\natexlab{a}})}]{evans2011reset}%
  \BibitemOpen
  \bibfield  {author} {\bibinfo {author} {\bibfnamefont {M.~R.}\ \bibnamefont
  {Evans}}\ and\ \bibinfo {author} {\bibfnamefont {S.~N.}\ \bibnamefont
  {Majumdar}},\ }\href {https://doi.org/10.1103/PhysRevLett.106.160601}
  {\bibfield  {journal} {\bibinfo  {journal} {Phys. Rev. Lett.}\ }\textbf
  {\bibinfo {volume} {106}},\ \bibinfo {pages} {160601} (\bibinfo {year}
  {2011}{\natexlab{a}})}\BibitemShut {NoStop}%
\bibitem [{\citenamefont {Evans}\ \emph {et~al.}(2020)\citenamefont {Evans},
  \citenamefont {Majumdar},\ and\ \citenamefont {Schehr}}]{evans2020review}%
  \BibitemOpen
  \bibfield  {author} {\bibinfo {author} {\bibfnamefont {M.~R.}\ \bibnamefont
  {Evans}}, \bibinfo {author} {\bibfnamefont {S.~N.}\ \bibnamefont
  {Majumdar}},\ and\ \bibinfo {author} {\bibfnamefont {G.}~\bibnamefont
  {Schehr}},\ }\href@noop {} {\bibfield  {journal} {\bibinfo  {journal} {J.
  Phys. A Math. Theor.}\ }\textbf {\bibinfo {volume} {53}},\ \bibinfo {pages}
  {193001} (\bibinfo {year} {2020})}\BibitemShut {NoStop}%
\bibitem [{\citenamefont {Bérut}\ \emph {et~al.}(2012)\citenamefont {Bérut},
  \citenamefont {Arakelyan}, \citenamefont {Petrosyan}, \citenamefont
  {Ciliberto}, \citenamefont {Dillenschneider},\ and\ \citenamefont
  {Lutz}}]{Berut2012MD}%
  \BibitemOpen
  \bibfield  {author} {\bibinfo {author} {\bibfnamefont {A.}~\bibnamefont
  {Bérut}}, \bibinfo {author} {\bibfnamefont {A.}~\bibnamefont {Arakelyan}},
  \bibinfo {author} {\bibfnamefont {A.}~\bibnamefont {Petrosyan}}, \bibinfo
  {author} {\bibfnamefont {S.}~\bibnamefont {Ciliberto}}, \bibinfo {author}
  {\bibfnamefont {R.}~\bibnamefont {Dillenschneider}},\ and\ \bibinfo {author}
  {\bibfnamefont {E.}~\bibnamefont {Lutz}},\ }\href@noop {} {\bibfield
  {journal} {\bibinfo  {journal} {Nature}\ }\textbf {\bibinfo {volume} {483}},\
  \bibinfo {pages} {187 } (\bibinfo {year} {2012})}\BibitemShut {NoStop}%
\bibitem [{\citenamefont {Koski}\ \emph {et~al.}(2014)\citenamefont {Koski},
  \citenamefont {Maisi}, \citenamefont {Pekola},\ and\ \citenamefont
  {Averin}}]{Koski2014szilard}%
  \BibitemOpen
  \bibfield  {author} {\bibinfo {author} {\bibfnamefont {J.~V.}\ \bibnamefont
  {Koski}}, \bibinfo {author} {\bibfnamefont {V.~F.}\ \bibnamefont {Maisi}},
  \bibinfo {author} {\bibfnamefont {J.~P.}\ \bibnamefont {Pekola}},\ and\
  \bibinfo {author} {\bibfnamefont {D.~V.}\ \bibnamefont {Averin}},\
  }\href@noop {} {\bibfield  {journal} {\bibinfo  {journal} {Proc. Natl. Acad.
  Sci. U.S.A.}\ }\textbf {\bibinfo {volume} {111}},\ \bibinfo {pages} {13786}
  (\bibinfo {year} {2014})}\BibitemShut {NoStop}%
\bibitem [{\citenamefont {Roldán}\ \emph {et~al.}(2014)\citenamefont
  {Roldán}, \citenamefont {Martínez}, \citenamefont {Parrondo},\ and\
  \citenamefont {Petrov}}]{Roldan2014symm}%
  \BibitemOpen
  \bibfield  {author} {\bibinfo {author} {\bibfnamefont {E.}~\bibnamefont
  {Roldán}}, \bibinfo {author} {\bibfnamefont {I.~A.}\ \bibnamefont
  {Martínez}}, \bibinfo {author} {\bibfnamefont {J.~M.~R.}\ \bibnamefont
  {Parrondo}},\ and\ \bibinfo {author} {\bibfnamefont {D.}~\bibnamefont
  {Petrov}},\ }\href@noop {} {\bibfield  {journal} {\bibinfo  {journal} {Nat.
  Phys.}\ }\textbf {\bibinfo {volume} {10}},\ \bibinfo {pages} {457} (\bibinfo
  {year} {2014})}\BibitemShut {NoStop}%
\bibitem [{\citenamefont {Rold\'an}\ \emph {et~al.}(2016)\citenamefont
  {Rold\'an}, \citenamefont {Lisica}, \citenamefont {S\'anchez-Taltavull},\
  and\ \citenamefont {Grill}}]{Roldan2016bio}%
  \BibitemOpen
  \bibfield  {author} {\bibinfo {author} {\bibfnamefont {E.}~\bibnamefont
  {Rold\'an}}, \bibinfo {author} {\bibfnamefont {A.}~\bibnamefont {Lisica}},
  \bibinfo {author} {\bibfnamefont {D.}~\bibnamefont {S\'anchez-Taltavull}},\
  and\ \bibinfo {author} {\bibfnamefont {S.~W.}\ \bibnamefont {Grill}},\ }\href
  {https://doi.org/10.1103/PhysRevE.93.062411} {\bibfield  {journal} {\bibinfo
  {journal} {Phys. Rev. E}\ }\textbf {\bibinfo {volume} {93}},\ \bibinfo
  {pages} {062411} (\bibinfo {year} {2016})}\BibitemShut {NoStop}%
\bibitem [{\citenamefont {Lisica}\ \emph {et~al.}(2016)\citenamefont {Lisica},
  \citenamefont {Engel}, \citenamefont {Jahnel}, \citenamefont {Édgar
  Roldán}, \citenamefont {Galburt}, \citenamefont {Cramer},\ and\
  \citenamefont {Grill}}]{Lisica2016back}%
  \BibitemOpen
  \bibfield  {author} {\bibinfo {author} {\bibfnamefont {A.}~\bibnamefont
  {Lisica}}, \bibinfo {author} {\bibfnamefont {C.}~\bibnamefont {Engel}},
  \bibinfo {author} {\bibfnamefont {M.}~\bibnamefont {Jahnel}}, \bibinfo
  {author} {\bibnamefont {Édgar Roldán}}, \bibinfo {author} {\bibfnamefont
  {E.~A.}\ \bibnamefont {Galburt}}, \bibinfo {author} {\bibfnamefont
  {P.}~\bibnamefont {Cramer}},\ and\ \bibinfo {author} {\bibfnamefont {S.~W.}\
  \bibnamefont {Grill}},\ }\href@noop {} {\bibfield  {journal} {\bibinfo
  {journal} {Proc. Natl. Acad. Sci. U.S.A.}\ }\textbf {\bibinfo {volume}
  {113}},\ \bibinfo {pages} {2946} (\bibinfo {year} {2016})}\BibitemShut
  {NoStop}%
\bibitem [{\citenamefont {Bressloff}(2020{\natexlab{a}})}]{Bressloff2020cyto}%
  \BibitemOpen
  \bibfield  {author} {\bibinfo {author} {\bibfnamefont {P.~C.}\ \bibnamefont
  {Bressloff}},\ }\href {https://doi.org/10.1088/1751-8121/ab7138} {\bibfield
  {journal} {\bibinfo  {journal} {J. Phys. A Math. Theor.}\ }\textbf {\bibinfo
  {volume} {53}},\ \bibinfo {pages} {105001} (\bibinfo {year}
  {2020}{\natexlab{a}})}\BibitemShut {NoStop}%
\bibitem [{\citenamefont {Bressloff}(2020{\natexlab{b}})}]{Bressloff2020intra}%
  \BibitemOpen
  \bibfield  {author} {\bibinfo {author} {\bibfnamefont {P.~C.}\ \bibnamefont
  {Bressloff}},\ }\href {https://doi.org/10.1088/1751-8121/ab9fb7} {\bibfield
  {journal} {\bibinfo  {journal} {J. Phys. A Math. Theor.}\ }\textbf {\bibinfo
  {volume} {53}},\ \bibinfo {pages} {355001} (\bibinfo {year}
  {2020}{\natexlab{b}})}\BibitemShut {NoStop}%
\bibitem [{\citenamefont {Genthon}\ \emph {et~al.}(2022)\citenamefont
  {Genthon}, \citenamefont {García-García},\ and\ \citenamefont
  {Lacoste}}]{Genthon2022cell}%
  \BibitemOpen
  \bibfield  {author} {\bibinfo {author} {\bibfnamefont {A.}~\bibnamefont
  {Genthon}}, \bibinfo {author} {\bibfnamefont {R.}~\bibnamefont
  {García-García}},\ and\ \bibinfo {author} {\bibfnamefont {D.}~\bibnamefont
  {Lacoste}},\ }\href {https://doi.org/10.1088/1751-8121/ac491a} {\bibfield
  {journal} {\bibinfo  {journal} {J. Phys. A Math. Theor.}\ }\textbf {\bibinfo
  {volume} {55}},\ \bibinfo {pages} {074001} (\bibinfo {year}
  {2022})}\BibitemShut {NoStop}%
\bibitem [{\citenamefont {Murashita}\ \emph {et~al.}(2014)\citenamefont
  {Murashita}, \citenamefont {Funo},\ and\ \citenamefont
  {Ueda}}]{Ueda2014_irrev}%
  \BibitemOpen
  \bibfield  {author} {\bibinfo {author} {\bibfnamefont {Y.}~\bibnamefont
  {Murashita}}, \bibinfo {author} {\bibfnamefont {K.}~\bibnamefont {Funo}},\
  and\ \bibinfo {author} {\bibfnamefont {M.}~\bibnamefont {Ueda}},\ }\href
  {https://doi.org/10.1103/PhysRevE.90.042110} {\bibfield  {journal} {\bibinfo
  {journal} {Phys. Rev. E}\ }\textbf {\bibinfo {volume} {90}},\ \bibinfo
  {pages} {042110} (\bibinfo {year} {2014})}\BibitemShut {NoStop}%
\bibitem [{\citenamefont {Busiello}\ \emph {et~al.}(2020)\citenamefont
  {Busiello}, \citenamefont {Gupta},\ and\ \citenamefont
  {Maritan}}]{Busiello2020uni}%
  \BibitemOpen
  \bibfield  {author} {\bibinfo {author} {\bibfnamefont {D.~M.}\ \bibnamefont
  {Busiello}}, \bibinfo {author} {\bibfnamefont {D.}~\bibnamefont {Gupta}},\
  and\ \bibinfo {author} {\bibfnamefont {A.}~\bibnamefont {Maritan}},\ }\href
  {https://doi.org/10.1103/PhysRevResearch.2.023011} {\bibfield  {journal}
  {\bibinfo  {journal} {Phys. Rev. Res.}\ }\textbf {\bibinfo {volume} {2}},\
  \bibinfo {pages} {023011} (\bibinfo {year} {2020})}\BibitemShut {NoStop}%
\bibitem [{\citenamefont {Fuchs}\ \emph {et~al.}(2016)\citenamefont {Fuchs},
  \citenamefont {Goldt},\ and\ \citenamefont {Seifert}}]{fuchs2016stochastic}%
  \BibitemOpen
  \bibfield  {author} {\bibinfo {author} {\bibfnamefont {J.}~\bibnamefont
  {Fuchs}}, \bibinfo {author} {\bibfnamefont {S.}~\bibnamefont {Goldt}},\ and\
  \bibinfo {author} {\bibfnamefont {U.}~\bibnamefont {Seifert}},\ }\href@noop
  {} {\bibfield  {journal} {\bibinfo  {journal} {Europhys. Lett.}\ }\textbf
  {\bibinfo {volume} {113}},\ \bibinfo {pages} {60009} (\bibinfo {year}
  {2016})}\BibitemShut {NoStop}%
\bibitem [{\citenamefont {Pal}\ and\ \citenamefont
  {Rahav}(2017)}]{pal2017integral}%
  \BibitemOpen
  \bibfield  {author} {\bibinfo {author} {\bibfnamefont {A.}~\bibnamefont
  {Pal}}\ and\ \bibinfo {author} {\bibfnamefont {S.}~\bibnamefont {Rahav}},\
  }\href@noop {} {\bibfield  {journal} {\bibinfo  {journal} {Phys. Rev. E}\
  }\textbf {\bibinfo {volume} {96}},\ \bibinfo {pages} {062135} (\bibinfo
  {year} {2017})}\BibitemShut {NoStop}%
\bibitem [{\citenamefont {Gupta}\ \emph
  {et~al.}(2020{\natexlab{a}})\citenamefont {Gupta}, \citenamefont {Plata},\
  and\ \citenamefont {Pal}}]{gupta2020work}%
  \BibitemOpen
  \bibfield  {author} {\bibinfo {author} {\bibfnamefont {D.}~\bibnamefont
  {Gupta}}, \bibinfo {author} {\bibfnamefont {C.~A.}\ \bibnamefont {Plata}},\
  and\ \bibinfo {author} {\bibfnamefont {A.}~\bibnamefont {Pal}},\ }\href@noop
  {} {\bibfield  {journal} {\bibinfo  {journal} {Phys. Rev. Lett.}\ }\textbf
  {\bibinfo {volume} {124}},\ \bibinfo {pages} {110608} (\bibinfo {year}
  {2020}{\natexlab{a}})}\BibitemShut {NoStop}%
\bibitem [{\citenamefont {Pal}\ \emph {et~al.}(2021)\citenamefont {Pal},
  \citenamefont {Reuveni},\ and\ \citenamefont {Rahav}}]{pal2021thermodynamic}%
  \BibitemOpen
  \bibfield  {author} {\bibinfo {author} {\bibfnamefont {A.}~\bibnamefont
  {Pal}}, \bibinfo {author} {\bibfnamefont {S.}~\bibnamefont {Reuveni}},\ and\
  \bibinfo {author} {\bibfnamefont {S.}~\bibnamefont {Rahav}},\ }\href@noop {}
  {\bibfield  {journal} {\bibinfo  {journal} {Phys. Rev. Res.}\ }\textbf
  {\bibinfo {volume} {3}},\ \bibinfo {pages} {013273} (\bibinfo {year}
  {2021})}\BibitemShut {NoStop}%
\bibitem [{\citenamefont {Seifert}(2005)}]{Seifert:2005epa}%
  \BibitemOpen
  \bibfield  {author} {\bibinfo {author} {\bibfnamefont {U.}~\bibnamefont
  {Seifert}},\ }\href@noop {} {\bibfield  {journal} {\bibinfo  {journal} {Phys.
  Rev. Lett.}\ }\textbf {\bibinfo {volume} {95}},\ \bibinfo {pages} {040602}
  (\bibinfo {year} {2005})}\BibitemShut {NoStop}%
\bibitem [{\citenamefont {Manrubia}\ and\ \citenamefont
  {Zanette}(1999)}]{Manrubia1999dis}%
  \BibitemOpen
  \bibfield  {author} {\bibinfo {author} {\bibfnamefont {S.~C.}\ \bibnamefont
  {Manrubia}}\ and\ \bibinfo {author} {\bibfnamefont {D.~H.}\ \bibnamefont
  {Zanette}},\ }\href@noop {} {\bibfield  {journal} {\bibinfo  {journal} {Phys.
  Rev. E}\ }\textbf {\bibinfo {volume} {59}},\ \bibinfo {pages} {4945}
  (\bibinfo {year} {1999})}\BibitemShut {NoStop}%
\bibitem [{\citenamefont {Evans}\ and\ \citenamefont
  {Majumdar}(2011{\natexlab{b}})}]{evans2011diffusion}%
  \BibitemOpen
  \bibfield  {author} {\bibinfo {author} {\bibfnamefont {M.~R.}\ \bibnamefont
  {Evans}}\ and\ \bibinfo {author} {\bibfnamefont {S.~N.}\ \bibnamefont
  {Majumdar}},\ }\href@noop {} {\bibfield  {journal} {\bibinfo  {journal} {J.
  Phys. A Math. Theor.}\ }\textbf {\bibinfo {volume} {44}},\ \bibinfo {pages}
  {435001} (\bibinfo {year} {2011}{\natexlab{b}})}\BibitemShut {NoStop}%
\bibitem [{\citenamefont {Toledo-Marin}\ and\ \citenamefont
  {Boyer}(2022)}]{toledo2022first}%
  \BibitemOpen
  \bibfield  {author} {\bibinfo {author} {\bibfnamefont {J.~Q.}\ \bibnamefont
  {Toledo-Marin}}\ and\ \bibinfo {author} {\bibfnamefont {D.}~\bibnamefont
  {Boyer}},\ }\href@noop {} {\bibfield  {journal} {\bibinfo  {journal}
  {preprint arXiv:2206.14387}\ } (\bibinfo {year} {2022})}\BibitemShut
  {NoStop}%
\bibitem [{\citenamefont {Gonz{\'a}lez}\ \emph {et~al.}(2021)\citenamefont
  {Gonz{\'a}lez}, \citenamefont {Riascos},\ and\ \citenamefont
  {Boyer}}]{gonzalez2021diffusive}%
  \BibitemOpen
  \bibfield  {author} {\bibinfo {author} {\bibfnamefont {F.~H.}\ \bibnamefont
  {Gonz{\'a}lez}}, \bibinfo {author} {\bibfnamefont {A.~P.}\ \bibnamefont
  {Riascos}},\ and\ \bibinfo {author} {\bibfnamefont {D.}~\bibnamefont
  {Boyer}},\ }\href@noop {} {\bibfield  {journal} {\bibinfo  {journal} {Phys.
  Rev. E}\ }\textbf {\bibinfo {volume} {103}},\ \bibinfo {pages} {062126}
  (\bibinfo {year} {2021})}\BibitemShut {NoStop}%
\bibitem [{\citenamefont {Dahlenburg}\ \emph {et~al.}(2021)\citenamefont
  {Dahlenburg}, \citenamefont {Chechkin}, \citenamefont {Schumer},\ and\
  \citenamefont {Metzler}}]{dahlenburg2021stochastic}%
  \BibitemOpen
  \bibfield  {author} {\bibinfo {author} {\bibfnamefont {M.}~\bibnamefont
  {Dahlenburg}}, \bibinfo {author} {\bibfnamefont {A.~V.}\ \bibnamefont
  {Chechkin}}, \bibinfo {author} {\bibfnamefont {R.}~\bibnamefont {Schumer}},\
  and\ \bibinfo {author} {\bibfnamefont {R.}~\bibnamefont {Metzler}},\
  }\href@noop {} {\bibfield  {journal} {\bibinfo  {journal} {Phys. Rev. E}\
  }\textbf {\bibinfo {volume} {103}},\ \bibinfo {pages} {052123} (\bibinfo
  {year} {2021})}\BibitemShut {NoStop}%
\bibitem [{\citenamefont {Besga}\ \emph {et~al.}(2020)\citenamefont {Besga},
  \citenamefont {Bovon}, \citenamefont {Petrosyan}, \citenamefont {Majumdar},\
  and\ \citenamefont {Ciliberto}}]{besga2020PRR}%
  \BibitemOpen
  \bibfield  {author} {\bibinfo {author} {\bibfnamefont {B.}~\bibnamefont
  {Besga}}, \bibinfo {author} {\bibfnamefont {A.}~\bibnamefont {Bovon}},
  \bibinfo {author} {\bibfnamefont {A.}~\bibnamefont {Petrosyan}}, \bibinfo
  {author} {\bibfnamefont {S.~N.}\ \bibnamefont {Majumdar}},\ and\ \bibinfo
  {author} {\bibfnamefont {S.}~\bibnamefont {Ciliberto}},\ }\href
  {https://doi.org/10.1103/PhysRevResearch.2.032029} {\bibfield  {journal}
  {\bibinfo  {journal} {Phys. Rev. Res.}\ }\textbf {\bibinfo {volume} {2}},\
  \bibinfo {pages} {032029} (\bibinfo {year} {2020})}\BibitemShut {NoStop}%
\bibitem [{\citenamefont {Faisant}\ \emph {et~al.}(2021)\citenamefont
  {Faisant}, \citenamefont {Besga}, \citenamefont {Petrosyan}, \citenamefont
  {Ciliberto},\ and\ \citenamefont {Majumdar}}]{faisant2021_2d}%
  \BibitemOpen
  \bibfield  {author} {\bibinfo {author} {\bibfnamefont {F.}~\bibnamefont
  {Faisant}}, \bibinfo {author} {\bibfnamefont {B.}~\bibnamefont {Besga}},
  \bibinfo {author} {\bibfnamefont {A.}~\bibnamefont {Petrosyan}}, \bibinfo
  {author} {\bibfnamefont {S.}~\bibnamefont {Ciliberto}},\ and\ \bibinfo
  {author} {\bibfnamefont {S.~N.}\ \bibnamefont {Majumdar}},\ }\href
  {https://doi.org/10.1088/1742-5468/ac2cc7} {\bibfield  {journal} {\bibinfo
  {journal} {J. Stat. Mech. Theory Exp.}\ }\textbf {\bibinfo {volume} {2021}},\
  \bibinfo {pages} {113203} (\bibinfo {year} {2021})}\BibitemShut {NoStop}%
\bibitem [{\citenamefont {Tal-Friedman}\ \emph {et~al.}(2020)\citenamefont
  {Tal-Friedman}, \citenamefont {Pal}, \citenamefont {Sekhon}, \citenamefont
  {Reuveni},\ and\ \citenamefont {Roichman}}]{friedman2020exp}%
  \BibitemOpen
  \bibfield  {author} {\bibinfo {author} {\bibfnamefont {O.}~\bibnamefont
  {Tal-Friedman}}, \bibinfo {author} {\bibfnamefont {A.}~\bibnamefont {Pal}},
  \bibinfo {author} {\bibfnamefont {A.}~\bibnamefont {Sekhon}}, \bibinfo
  {author} {\bibfnamefont {S.}~\bibnamefont {Reuveni}},\ and\ \bibinfo {author}
  {\bibfnamefont {Y.}~\bibnamefont {Roichman}},\ }\href@noop {} {\bibfield
  {journal} {\bibinfo  {journal} {J. Phys. Chem. Lett.}\ }\textbf {\bibinfo
  {volume} {11}},\ \bibinfo {pages} {7350} (\bibinfo {year}
  {2020})}\BibitemShut {NoStop}%
\bibitem [{\citenamefont {Gupta}\ and\ \citenamefont
  {Plata}(2022)}]{Deepak2022_work}%
  \BibitemOpen
  \bibfield  {author} {\bibinfo {author} {\bibfnamefont {D.}~\bibnamefont
  {Gupta}}\ and\ \bibinfo {author} {\bibfnamefont {C.~A.}\ \bibnamefont
  {Plata}},\ }\href@noop {} {\bibfield  {journal} {\bibinfo  {journal} {New J.
  Phys. in press https://doi.org/10.1088/1367-2630/aca25e}\ } (\bibinfo {year}
  {2022})}\BibitemShut {NoStop}%
\bibitem [{\citenamefont {Evans}\ and\ \citenamefont
  {Majumdar}(2011{\natexlab{c}})}]{evans2011optimal}%
  \BibitemOpen
  \bibfield  {author} {\bibinfo {author} {\bibfnamefont {M.~R.}\ \bibnamefont
  {Evans}}\ and\ \bibinfo {author} {\bibfnamefont {S.~N.}\ \bibnamefont
  {Majumdar}},\ }\href {https://doi.org/10.1088/1751-8113/44/43/435001}
  {\bibfield  {journal} {\bibinfo  {journal} {J. Phys. A Math. Theor.}\
  }\textbf {\bibinfo {volume} {44}},\ \bibinfo {pages} {435001} (\bibinfo
  {year} {2011}{\natexlab{c}})}\BibitemShut {NoStop}%
\bibitem [{\citenamefont {Seifert}(2012)}]{Seifert:2012stf}%
  \BibitemOpen
  \bibfield  {author} {\bibinfo {author} {\bibfnamefont {U.}~\bibnamefont
  {Seifert}},\ }\href {http://stacks.iop.org/0034-4885/75/i=12/a=126001}
  {\bibfield  {journal} {\bibinfo  {journal} {Rep. Prog. Phys.}\ }\textbf
  {\bibinfo {volume} {75}},\ \bibinfo {pages} {126001} (\bibinfo {year}
  {2012})}\BibitemShut {NoStop}%
\bibitem [{\citenamefont {Qian}(2001)}]{qian2001mesoscopic}%
  \BibitemOpen
  \bibfield  {author} {\bibinfo {author} {\bibfnamefont {H.}~\bibnamefont
  {Qian}},\ }\href@noop {} {\bibfield  {journal} {\bibinfo  {journal} {Phys.
  Rev. E}\ }\textbf {\bibinfo {volume} {65}},\ \bibinfo {pages} {016102}
  (\bibinfo {year} {2001})}\BibitemShut {NoStop}%
\bibitem [{\citenamefont {Busiello}\ \emph {et~al.}(2019)\citenamefont
  {Busiello}, \citenamefont {Hidalgo},\ and\ \citenamefont
  {Maritan}}]{Busiello2019coarse}%
  \BibitemOpen
  \bibfield  {author} {\bibinfo {author} {\bibfnamefont {D.~M.}\ \bibnamefont
  {Busiello}}, \bibinfo {author} {\bibfnamefont {J.}~\bibnamefont {Hidalgo}},\
  and\ \bibinfo {author} {\bibfnamefont {A.}~\bibnamefont {Maritan}},\ }\href
  {https://doi.org/10.1088/1367-2630/ab29c0} {\bibfield  {journal} {\bibinfo
  {journal} {New J. Phys.}\ }\textbf {\bibinfo {volume} {21}},\ \bibinfo
  {pages} {073004} (\bibinfo {year} {2019})}\BibitemShut {NoStop}%
\bibitem [{\citenamefont {Van~den Broeck}\ and\ \citenamefont
  {Esposito}(2010)}]{broeck2010FP}%
  \BibitemOpen
  \bibfield  {author} {\bibinfo {author} {\bibfnamefont {C.}~\bibnamefont
  {Van~den Broeck}}\ and\ \bibinfo {author} {\bibfnamefont {M.}~\bibnamefont
  {Esposito}},\ }\href {https://doi.org/10.1103/PhysRevE.82.011144} {\bibfield
  {journal} {\bibinfo  {journal} {Phys. Rev. E}\ }\textbf {\bibinfo {volume}
  {82}},\ \bibinfo {pages} {011144} (\bibinfo {year} {2010})}\BibitemShut
  {NoStop}%
\bibitem [{\citenamefont {Meylahn}\ \emph {et~al.}(2015)\citenamefont
  {Meylahn}, \citenamefont {Sabhapandit},\ and\ \citenamefont
  {Touchette}}]{meylahn2015LD}%
  \BibitemOpen
  \bibfield  {author} {\bibinfo {author} {\bibfnamefont {J.~M.}\ \bibnamefont
  {Meylahn}}, \bibinfo {author} {\bibfnamefont {S.}~\bibnamefont
  {Sabhapandit}},\ and\ \bibinfo {author} {\bibfnamefont {H.}~\bibnamefont
  {Touchette}},\ }\href {https://doi.org/10.1103/PhysRevE.92.062148} {\bibfield
   {journal} {\bibinfo  {journal} {Phys. Rev. E}\ }\textbf {\bibinfo {volume}
  {92}},\ \bibinfo {pages} {062148} (\bibinfo {year} {2015})}\BibitemShut
  {NoStop}%
\bibitem [{\citenamefont {Harris}\ and\ \citenamefont
  {Touchette}(2017)}]{harris2017phase}%
  \BibitemOpen
  \bibfield  {author} {\bibinfo {author} {\bibfnamefont {R.~J.}\ \bibnamefont
  {Harris}}\ and\ \bibinfo {author} {\bibfnamefont {H.}~\bibnamefont
  {Touchette}},\ }\href@noop {} {\bibfield  {journal} {\bibinfo  {journal} {J.
  Phys. A Math. Theor.}\ }\textbf {\bibinfo {volume} {50}},\ \bibinfo {pages}
  {10LT01} (\bibinfo {year} {2017})}\BibitemShut {NoStop}%
\bibitem [{\citenamefont {Coghi}\ and\ \citenamefont
  {Harris}(2020)}]{coghi2020large}%
  \BibitemOpen
  \bibfield  {author} {\bibinfo {author} {\bibfnamefont {F.}~\bibnamefont
  {Coghi}}\ and\ \bibinfo {author} {\bibfnamefont {R.~J.}\ \bibnamefont
  {Harris}},\ }\href@noop {} {\bibfield  {journal} {\bibinfo  {journal} {J.
  Stat. Phys.}\ }\textbf {\bibinfo {volume} {179}},\ \bibinfo {pages} {131}
  (\bibinfo {year} {2020})}\BibitemShut {NoStop}%
\bibitem [{\citenamefont {Smith}\ and\ \citenamefont {Majumdar}(2022)}]{SM22}%
  \BibitemOpen
  \bibfield  {author} {\bibinfo {author} {\bibfnamefont {N.~R.}\ \bibnamefont
  {Smith}}\ and\ \bibinfo {author} {\bibfnamefont {S.~N.}\ \bibnamefont
  {Majumdar}},\ }\href@noop {} {\bibfield  {journal} {\bibinfo  {journal} {J.
  Stat. Mech. Theory Exp.}\ }\textbf {\bibinfo {volume} {2022}},\ \bibinfo
  {pages} {053212} (\bibinfo {year} {2022})}\BibitemShut {NoStop}%
\bibitem [{\citenamefont {Mori}\ \emph
  {et~al.}(2021{\natexlab{a}})\citenamefont {Mori}, \citenamefont {Le~Doussal},
  \citenamefont {Majumdar},\ and\ \citenamefont {Schehr}}]{MLDM21}%
  \BibitemOpen
  \bibfield  {author} {\bibinfo {author} {\bibfnamefont {F.}~\bibnamefont
  {Mori}}, \bibinfo {author} {\bibfnamefont {P.}~\bibnamefont {Le~Doussal}},
  \bibinfo {author} {\bibfnamefont {S.~N.}\ \bibnamefont {Majumdar}},\ and\
  \bibinfo {author} {\bibfnamefont {G.}~\bibnamefont {Schehr}},\ }\href@noop {}
  {\bibfield  {journal} {\bibinfo  {journal} {Phys. Rev. E}\ }\textbf {\bibinfo
  {volume} {103}},\ \bibinfo {pages} {062134} (\bibinfo {year}
  {2021}{\natexlab{a}})}\BibitemShut {NoStop}%
\bibitem [{\citenamefont {Majumdar}\ \emph {et~al.}(2005)\citenamefont
  {Majumdar}, \citenamefont {Evans},\ and\ \citenamefont
  {Zia}}]{majumdar2005nature}%
  \BibitemOpen
  \bibfield  {author} {\bibinfo {author} {\bibfnamefont {S.~N.}\ \bibnamefont
  {Majumdar}}, \bibinfo {author} {\bibfnamefont {M.}~\bibnamefont {Evans}},\
  and\ \bibinfo {author} {\bibfnamefont {R.}~\bibnamefont {Zia}},\ }\href@noop
  {} {\bibfield  {journal} {\bibinfo  {journal} {Phys. Rev. Lett.}\ }\textbf
  {\bibinfo {volume} {94}},\ \bibinfo {pages} {180601} (\bibinfo {year}
  {2005})}\BibitemShut {NoStop}%
\bibitem [{\citenamefont {Filiasi}\ \emph {et~al.}(2014)\citenamefont
  {Filiasi}, \citenamefont {Livan}, \citenamefont {Marsili}, \citenamefont
  {Peressi}, \citenamefont {Vesselli},\ and\ \citenamefont
  {Zarinelli}}]{filiasi2014concentration}%
  \BibitemOpen
  \bibfield  {author} {\bibinfo {author} {\bibfnamefont {M.}~\bibnamefont
  {Filiasi}}, \bibinfo {author} {\bibfnamefont {G.}~\bibnamefont {Livan}},
  \bibinfo {author} {\bibfnamefont {M.}~\bibnamefont {Marsili}}, \bibinfo
  {author} {\bibfnamefont {M.}~\bibnamefont {Peressi}}, \bibinfo {author}
  {\bibfnamefont {E.}~\bibnamefont {Vesselli}},\ and\ \bibinfo {author}
  {\bibfnamefont {E.}~\bibnamefont {Zarinelli}},\ }\href@noop {} {\bibfield
  {journal} {\bibinfo  {journal} {J. Stat. Mech. Theory Exp.}\ }\textbf
  {\bibinfo {volume} {2014}},\ \bibinfo {pages} {P09030} (\bibinfo {year}
  {2014})}\BibitemShut {NoStop}%
\bibitem [{\citenamefont {Gradenigo}\ and\ \citenamefont
  {Majumdar}(2019)}]{GM19}%
  \BibitemOpen
  \bibfield  {author} {\bibinfo {author} {\bibfnamefont {G.}~\bibnamefont
  {Gradenigo}}\ and\ \bibinfo {author} {\bibfnamefont {S.~N.}\ \bibnamefont
  {Majumdar}},\ }\href@noop {} {\bibfield  {journal} {\bibinfo  {journal} {J.
  Stat. Mech. Theory Exp.}\ }\textbf {\bibinfo {volume} {2019}},\ \bibinfo
  {pages} {053206} (\bibinfo {year} {2019})}\BibitemShut {NoStop}%
\bibitem [{\citenamefont {Mori}\ \emph
  {et~al.}(2021{\natexlab{b}})\citenamefont {Mori}, \citenamefont {Gradenigo},\
  and\ \citenamefont {Majumdar}}]{MGM21}%
  \BibitemOpen
  \bibfield  {author} {\bibinfo {author} {\bibfnamefont {F.}~\bibnamefont
  {Mori}}, \bibinfo {author} {\bibfnamefont {G.}~\bibnamefont {Gradenigo}},\
  and\ \bibinfo {author} {\bibfnamefont {S.~N.}\ \bibnamefont {Majumdar}},\
  }\href@noop {} {\bibfield  {journal} {\bibinfo  {journal} {J. Stat. Mech.
  Theory Exp.}\ }\textbf {\bibinfo {volume} {2021}},\ \bibinfo {pages} {103208}
  (\bibinfo {year} {2021}{\natexlab{b}})}\BibitemShut {NoStop}%
\bibitem [{\citenamefont {Gradenigo}\ \emph {et~al.}(2021)\citenamefont
  {Gradenigo}, \citenamefont {Iubini}, \citenamefont {Livi},\ and\
  \citenamefont {Majumdar}}]{gradenigo2021condensation}%
  \BibitemOpen
  \bibfield  {author} {\bibinfo {author} {\bibfnamefont {G.}~\bibnamefont
  {Gradenigo}}, \bibinfo {author} {\bibfnamefont {S.}~\bibnamefont {Iubini}},
  \bibinfo {author} {\bibfnamefont {R.}~\bibnamefont {Livi}},\ and\ \bibinfo
  {author} {\bibfnamefont {S.~N.}\ \bibnamefont {Majumdar}},\ }\href@noop {}
  {\bibfield  {journal} {\bibinfo  {journal} {Eur. Phys. J. E}\ }\textbf
  {\bibinfo {volume} {44}},\ \bibinfo {pages} {1} (\bibinfo {year}
  {2021})}\BibitemShut {NoStop}%
\bibitem [{\citenamefont {Gallavotti}\ and\ \citenamefont
  {Cohen}(1995)}]{gallavotti1995dynamical}%
  \BibitemOpen
  \bibfield  {author} {\bibinfo {author} {\bibfnamefont {G.}~\bibnamefont
  {Gallavotti}}\ and\ \bibinfo {author} {\bibfnamefont {E.~G.~D.}\ \bibnamefont
  {Cohen}},\ }\href@noop {} {\bibfield  {journal} {\bibinfo  {journal} {Phys.
  Rev. Lett.}\ }\textbf {\bibinfo {volume} {74}},\ \bibinfo {pages} {2694}
  (\bibinfo {year} {1995})}\BibitemShut {NoStop}%
\bibitem [{\citenamefont {Touchette}(2009)}]{Touchette:2009lda}%
  \BibitemOpen
  \bibfield  {author} {\bibinfo {author} {\bibfnamefont {H.}~\bibnamefont
  {Touchette}},\ }\href
  {https://doi.org/https://doi.org/10.1016/j.physrep.2009.05.002} {\bibfield
  {journal} {\bibinfo  {journal} {Phys. Rep.}\ }\textbf {\bibinfo {volume}
  {478}},\ \bibinfo {pages} {1 } (\bibinfo {year} {2009})}\BibitemShut
  {NoStop}%
\bibitem [{\citenamefont {Nagaev}(1969)}]{nagaev1969integral}%
  \BibitemOpen
  \bibfield  {author} {\bibinfo {author} {\bibfnamefont {A.~V.}\ \bibnamefont
  {Nagaev}},\ }\href@noop {} {\bibfield  {journal} {\bibinfo  {journal} {Theory
  of Probability \& Its Applications}\ }\textbf {\bibinfo {volume} {14}},\
  \bibinfo {pages} {51} (\bibinfo {year} {1969})}\BibitemShut {NoStop}%
\bibitem [{\citenamefont {Brosset}\ \emph {et~al.}(2020)\citenamefont
  {Brosset}, \citenamefont {Klein}, \citenamefont {Lagnoux},\ and\
  \citenamefont {Petit}}]{brosset2020probabilistic}%
  \BibitemOpen
  \bibfield  {author} {\bibinfo {author} {\bibfnamefont {F.}~\bibnamefont
  {Brosset}}, \bibinfo {author} {\bibfnamefont {T.}~\bibnamefont {Klein}},
  \bibinfo {author} {\bibfnamefont {A.}~\bibnamefont {Lagnoux}},\ and\ \bibinfo
  {author} {\bibfnamefont {P.}~\bibnamefont {Petit}},\ }\href@noop {}
  {\bibfield  {journal} {\bibinfo  {journal} {preprint arXiv:2007.08164}\ }
  (\bibinfo {year} {2020})}\BibitemShut {NoStop}%
\bibitem [{\citenamefont {De~Bruyne}\ and\ \citenamefont
  {Mori}(2021)}]{de2021resetting}%
  \BibitemOpen
  \bibfield  {author} {\bibinfo {author} {\bibfnamefont {B.}~\bibnamefont
  {De~Bruyne}}\ and\ \bibinfo {author} {\bibfnamefont {F.}~\bibnamefont
  {Mori}},\ }\href@noop {} {\bibfield  {journal} {\bibinfo  {journal} {preprint
  arXiv:2112.11416}\ } (\bibinfo {year} {2021})}\BibitemShut {NoStop}%
\bibitem [{\citenamefont {Gupta}\ \emph
  {et~al.}(2020{\natexlab{b}})\citenamefont {Gupta}, \citenamefont {Plata},
  \citenamefont {Kundu},\ and\ \citenamefont {Pal}}]{Gupta2021_finite}%
  \BibitemOpen
  \bibfield  {author} {\bibinfo {author} {\bibfnamefont {D.}~\bibnamefont
  {Gupta}}, \bibinfo {author} {\bibfnamefont {C.~A.}\ \bibnamefont {Plata}},
  \bibinfo {author} {\bibfnamefont {A.}~\bibnamefont {Kundu}},\ and\ \bibinfo
  {author} {\bibfnamefont {A.}~\bibnamefont {Pal}},\ }\href
  {https://doi.org/10.1088/1751-8121/abcf0b} {\bibfield  {journal} {\bibinfo
  {journal} {J. Phys. A Math. Theor.}\ }\textbf {\bibinfo {volume} {54}},\
  \bibinfo {pages} {025003} (\bibinfo {year} {2020}{\natexlab{b}})}\BibitemShut
  {NoStop}%
\bibitem [{\citenamefont {Pal}\ \emph {et~al.}(2019)\citenamefont {Pal},
  \citenamefont {Ku\ifmmode~\acute{s}\else \'{s}\fi{}mierz},\ and\
  \citenamefont {Reuveni}}]{pal2019_finite}%
  \BibitemOpen
  \bibfield  {author} {\bibinfo {author} {\bibfnamefont {A.}~\bibnamefont
  {Pal}}, \bibinfo {author} {\bibfnamefont {L.}~\bibnamefont
  {Ku\ifmmode~\acute{s}\else \'{s}\fi{}mierz}},\ and\ \bibinfo {author}
  {\bibfnamefont {S.}~\bibnamefont {Reuveni}},\ }\href
  {https://doi.org/10.1103/PhysRevE.100.040101} {\bibfield  {journal} {\bibinfo
   {journal} {Phys. Rev. E}\ }\textbf {\bibinfo {volume} {100}},\ \bibinfo
  {pages} {040101} (\bibinfo {year} {2019})}\BibitemShut {NoStop}%
\bibitem [{\citenamefont {Mas{\'o}-Puigdellosas}\ \emph
  {et~al.}(2019)\citenamefont {Mas{\'o}-Puigdellosas}, \citenamefont {Campos},\
  and\ \citenamefont {M{\'e}ndez}}]{maso2019transport}%
  \BibitemOpen
  \bibfield  {author} {\bibinfo {author} {\bibfnamefont {A.}~\bibnamefont
  {Mas{\'o}-Puigdellosas}}, \bibinfo {author} {\bibfnamefont {D.}~\bibnamefont
  {Campos}},\ and\ \bibinfo {author} {\bibfnamefont {V.}~\bibnamefont
  {M{\'e}ndez}},\ }\href@noop {} {\bibfield  {journal} {\bibinfo  {journal}
  {Phys. Rev. E}\ }\textbf {\bibinfo {volume} {100}},\ \bibinfo {pages}
  {042104} (\bibinfo {year} {2019})}\BibitemShut {NoStop}%
\bibitem [{\citenamefont {Radice}(2022)}]{radice2022diffusion}%
  \BibitemOpen
  \bibfield  {author} {\bibinfo {author} {\bibfnamefont {M.}~\bibnamefont
  {Radice}},\ }\href@noop {} {\bibfield  {journal} {\bibinfo  {journal} {J.
  Phys. A Math. Theor.}\ }\textbf {\bibinfo {volume} {55}},\ \bibinfo {pages}
  {224002} (\bibinfo {year} {2022})}\BibitemShut {NoStop}%
\bibitem [{\citenamefont {Reuveni}(2016)}]{reuveni2016optimal}%
  \BibitemOpen
  \bibfield  {author} {\bibinfo {author} {\bibfnamefont {S.}~\bibnamefont
  {Reuveni}},\ }\href@noop {} {\bibfield  {journal} {\bibinfo  {journal} {Phys.
  Rev. Lett.}\ }\textbf {\bibinfo {volume} {116}},\ \bibinfo {pages} {170601}
  (\bibinfo {year} {2016})}\BibitemShut {NoStop}%
\bibitem [{\citenamefont {Mercado-V{\'a}squez}\ \emph
  {et~al.}(2020)\citenamefont {Mercado-V{\'a}squez}, \citenamefont {Boyer},
  \citenamefont {Majumdar},\ and\ \citenamefont
  {Schehr}}]{mercado2020intermittent}%
  \BibitemOpen
  \bibfield  {author} {\bibinfo {author} {\bibfnamefont {G.}~\bibnamefont
  {Mercado-V{\'a}squez}}, \bibinfo {author} {\bibfnamefont {D.}~\bibnamefont
  {Boyer}}, \bibinfo {author} {\bibfnamefont {S.~N.}\ \bibnamefont
  {Majumdar}},\ and\ \bibinfo {author} {\bibfnamefont {G.}~\bibnamefont
  {Schehr}},\ }\href@noop {} {\bibfield  {journal} {\bibinfo  {journal} {J.
  Stat. Mech. Theory Exp.}\ }\textbf {\bibinfo {volume} {2020}},\ \bibinfo
  {pages} {113203} (\bibinfo {year} {2020})}\BibitemShut {NoStop}%
\bibitem [{\citenamefont {Mercado-V{\'a}squez}\ \emph
  {et~al.}(2022)\citenamefont {Mercado-V{\'a}squez}, \citenamefont {Boyer},\
  and\ \citenamefont {Majumdar}}]{mercado2022reducing}%
  \BibitemOpen
  \bibfield  {author} {\bibinfo {author} {\bibfnamefont {G.}~\bibnamefont
  {Mercado-V{\'a}squez}}, \bibinfo {author} {\bibfnamefont {D.}~\bibnamefont
  {Boyer}},\ and\ \bibinfo {author} {\bibfnamefont {S.~N.}\ \bibnamefont
  {Majumdar}},\ }\href@noop {} {\bibfield  {journal} {\bibinfo  {journal} {J.
  Stat. Mech. Theory Exp.}\ }\textbf {\bibinfo {volume} {2022}},\ \bibinfo
  {pages} {093202} (\bibinfo {year} {2022})}\BibitemShut {NoStop}%
\bibitem [{\citenamefont {Santra}\ \emph {et~al.}(2021)\citenamefont {Santra},
  \citenamefont {Das},\ and\ \citenamefont {Nath}}]{santra2021brownian}%
  \BibitemOpen
  \bibfield  {author} {\bibinfo {author} {\bibfnamefont {I.}~\bibnamefont
  {Santra}}, \bibinfo {author} {\bibfnamefont {S.}~\bibnamefont {Das}},\ and\
  \bibinfo {author} {\bibfnamefont {S.~K.}\ \bibnamefont {Nath}},\ }\href@noop
  {} {\bibfield  {journal} {\bibinfo  {journal} {J. Phys. A Math. Theor.}\
  }\textbf {\bibinfo {volume} {54}},\ \bibinfo {pages} {334001} (\bibinfo
  {year} {2021})}\BibitemShut {NoStop}%
\bibitem [{\citenamefont {Gupta}\ \emph
  {et~al.}(2020{\natexlab{c}})\citenamefont {Gupta}, \citenamefont {Plata},
  \citenamefont {Kundu},\ and\ \citenamefont {Pal}}]{gupta2020stochastic}%
  \BibitemOpen
  \bibfield  {author} {\bibinfo {author} {\bibfnamefont {D.}~\bibnamefont
  {Gupta}}, \bibinfo {author} {\bibfnamefont {C.~A.}\ \bibnamefont {Plata}},
  \bibinfo {author} {\bibfnamefont {A.}~\bibnamefont {Kundu}},\ and\ \bibinfo
  {author} {\bibfnamefont {A.}~\bibnamefont {Pal}},\ }\href@noop {} {\bibfield
  {journal} {\bibinfo  {journal} {J. Phys. A Math. Theor.}\ }\textbf {\bibinfo
  {volume} {54}},\ \bibinfo {pages} {025003} (\bibinfo {year}
  {2020}{\natexlab{c}})}\BibitemShut {NoStop}%
\bibitem [{\citenamefont {Gupta}\ \emph {et~al.}(2021)\citenamefont {Gupta},
  \citenamefont {Pal},\ and\ \citenamefont {Kundu}}]{gupta2021resetting}%
  \BibitemOpen
  \bibfield  {author} {\bibinfo {author} {\bibfnamefont {D.}~\bibnamefont
  {Gupta}}, \bibinfo {author} {\bibfnamefont {A.}~\bibnamefont {Pal}},\ and\
  \bibinfo {author} {\bibfnamefont {A.}~\bibnamefont {Kundu}},\ }\href@noop {}
  {\bibfield  {journal} {\bibinfo  {journal} {J. Stat. Mech. Theory Exp.}\
  }\textbf {\bibinfo {volume} {2021}},\ \bibinfo {pages} {043202} (\bibinfo
  {year} {2021})}\BibitemShut {NoStop}%
\bibitem [{\citenamefont {Alston}\ \emph {et~al.}(2022)\citenamefont {Alston},
  \citenamefont {Cocconi},\ and\ \citenamefont {Bertrand}}]{alston2022non}%
  \BibitemOpen
  \bibfield  {author} {\bibinfo {author} {\bibfnamefont {H.}~\bibnamefont
  {Alston}}, \bibinfo {author} {\bibfnamefont {L.}~\bibnamefont {Cocconi}},\
  and\ \bibinfo {author} {\bibfnamefont {T.}~\bibnamefont {Bertrand}},\
  }\href@noop {} {\bibfield  {journal} {\bibinfo  {journal} {J. Phys. A Math.
  Theor.}\ }\textbf {\bibinfo {volume} {55}},\ \bibinfo {pages} {274004}
  (\bibinfo {year} {2022})}\BibitemShut {NoStop}%
\bibitem [{\citenamefont {Schnakenberg}(1976)}]{schnakenberg1976network}%
  \BibitemOpen
  \bibfield  {author} {\bibinfo {author} {\bibfnamefont {J.}~\bibnamefont
  {Schnakenberg}},\ }\href@noop {} {\bibfield  {journal} {\bibinfo  {journal}
  {Rev. Mod. Phys.}\ }\textbf {\bibinfo {volume} {48}},\ \bibinfo {pages} {571}
  (\bibinfo {year} {1976})}\BibitemShut {NoStop}%
\end{thebibliography}%

\end{document}